\documentclass[aip,graphicx]{revtex4-2}

\usepackage{amsmath}
\usepackage{graphicx}
\usepackage{color}

\begin{document}
\title{Theoretical modeling of the dynamic range of an elastic nanobeam under tension with a geometric nonlinearity}

\author{N. W. Welles}
\affiliation{Department of Mechanical Engineering, Virginia Tech, Blacksburg, Virginia 24061, USA}

\author{M. Ma}
\affiliation{Department of Mechanical Engineering, Division of Materials Science and Engineering, and the Photonics Center, Boston University, Boston, Massachusetts 02215, USA}

\author{K. L. Ekinci}
\affiliation{Department of Mechanical Engineering, Division of Materials Science and Engineering, and the Photonics Center, Boston University, Boston, Massachusetts 02215, USA}

\author{M. R. Paul}
\email{mrp@vt.edu}
\affiliation{Department of Mechanical Engineering, Virginia Tech, Blacksburg, Virginia 24061, USA}

\date{\today}

\begin{abstract}
A theoretical description of the weakly nonlinear and mode-dependent dynamics of a nanoscale beam that is under intrinsic tension is developed. A full analysis of the dynamic range of the beam over a wide range of conditions is presented. The dynamic range is bounded from below by the amplitude of vibration due to thermal motion and it is bounded from above by large amplitude oscillations where the geometric nonlinearity plays a significant role due to stretching induced tension. The dynamics are analyzed using a beam with clamped boundaries, a string model, and a beam with hinged boundaries. The range of validity for the different models is quantified in detail. A hinged beam model is found to provide an accurate description, with insightful closed-form analytical expressions, over a wide range of conditions. The relative importance of bending and tension in the mode-dependent dynamics of the beam is determined. Bending is shown to be important for the higher modes of oscillation with the onset of its importance dependent upon the amount of intrinsic tension that is present. The theoretical predictions are directly compared with experimental measurements for the first ten modes of two nanoscale beams. We discuss the accuracy of these approaches and their use for the development of emerging micro and nanoscale technologies that exploit the multimodal dynamics of small elastic beams operating in the linear regime. 
\end{abstract}

\maketitle

\section{Introduction}

At the heart of many emerging micro and nanoscale technologies with increasing measurement sensitivities~\cite{ekinci:2005,bachtold:2022} are oscillating mechanical structures that are driven to the limits of their linear behavior to operate near the regime where nonlinearity becomes significant~\cite{lifshitz:2008,ma:2024,xu:2022,dinh:2023}. The linear dynamic range of an oscillator is an insightful quantity that describes the conditions for which the dynamics can be treated linearly in the sense that the amplitude of oscillation is proportional to the magnitude of the driving force. An understanding of the range of conditions yielding linear dynamics is a crucial element in the development of the operation of many technologies. This has yielded force sensors with zeptonewton  resolution~\cite{moser:2013,debonis:2018}, mass sensing at the yoctogram scale~\cite{jensen:2008,chaste:2012,hanay:2015}, and rheometers probing nanoliters of fluid~\cite{boskovic:2002,radiom:2012,waigh:2016}, to name a few.

In many technologies, the mechanical oscillator is a doubly-clamped elastic structure such as a nanobeam~\cite{kozinsky:2007,kouh:2017,barbish:2022,gress:2023,ma:2024}, nanowire~\cite{husain:2003,feng:2007}, nanotube~\cite{moser:2013,barnard:2019}, or suspended graphene sheet~\cite{bunch:2007,dinh:2023}. The idea of the linear dynamic range applies to the fundamental mode as well as the higher modes of oscillation. The higher modes of oscillation continue to draw interest for a variety of applications due to their increased natural frequencies~\cite{bargatin:2007,barbish:2022,gress:2023,ma:2024}.

The nonlinear dynamics of micro and nanoscale elastic structures now has a rich literature~\cite{lifshitz:2008,bachtold:2022,welles:2024}. For example, in one of the first measurements to experimentally probe hysteresis, Husain~\emph{et al.}~\cite{husain:2003} explored the linear and nonlinear dynamics of a magnetically driven platinum nanowire. Postma~\emph{et al.}~\cite{postma:2005} theoretically investigated the linear dynamic range of the fundamental mode of high aspect-ratio resonators including nanobeams, nanotubes, and nanowires. Using the mode shape of a buckled beam as an approximation for the fundamental mode shape of a doubly clamped beam under tension, a very strong dependence of the linear dynamic range on the aspect ratio was found. Kozinsky~\emph{et al.}~\cite{kozinsky:2006} demonstrated that the linear dynamic range of the fundamental mode of a nanobeam can be extended by electrostatically tuning the nonlinearity. Lifshitz and Cross~\cite{lifshitz:2008} provided an insightful theoretical investigation into the linear dynamic range of small beams using the Euler-Bernoulli beam and string approximations for the fundamental mode.

Recently, an investigation by Ma \emph{et al.}~\cite{ma:2024} experimentally measured the mode-dependent linear dynamic range of two SiN nanobeams that were under significant tension. The linear dynamic range of both beams was found to increase gradually with increasing mode number. In an effort to capture these measured trends for beams under tension, a string model was used which included tension and, by definition, did not include contributions from bending. The string model provided very good agreement with the experimentally measured beam with the highest tension, and it was found that the agreement decreased for the beam with a lower amount of tension. The relative roles of bending and tension with respect to the linear dynamic range was not addressed and was left as an open question. 

In this paper, we conduct a thorough theoretical investigation of the linear dynamic range of nanobeams for arbitrary values of the tension. We specifically focus upon the role of bending and tension on the mode-dependent linear dynamic range. In our theoretical development we present results for three different models. First, we solve the problem using the full equations of elasticity starting from the Euler-Bernoulli beam equation with clamped boundaries. We next provide the full and complete analysis for a string model. Finally, we present a theoretical description using the Euler-Bernoulli beam equation with a hinged boundary condition that is particularly useful in the presence of tension and an accurate closed-form description is desired that includes contributions from tension and bending. 

The remainder of the paper is organized as follows. In \S\ref{section:discussion} the essential physical ideas to describe the dynamics of a nanobeam are introduced. We describe the important roles of intrinsic tension, stretching induced tension, and bending and discuss how these influence the externally driven and thermally driven dynamics of the nanobeam. In \S\ref{section:ldr} the  expressions are developed for the mode-dependent linear dynamic range for a beam using three different models: a doubly-clamped beam, a string, and a hinged-beam.  In this section, the hinged-beam is used to present new insights characterizing the relative influence of bending and tension with increasing mode number. In \S\ref{section:comparison} we compare our theoretical descriptions directly with experimental measurement. Lastly, in \S\ref{section:conclusion} concluding remarks are presented.

\section{Discussion}
\label{section:discussion}

\subsection{The Euler-Bernoulli beam with intrinsic tension and a geometric nonlinearity}

\begin{figure}[h!]
\vspace{1cm}
\begin{center}
\includegraphics[width=5in]{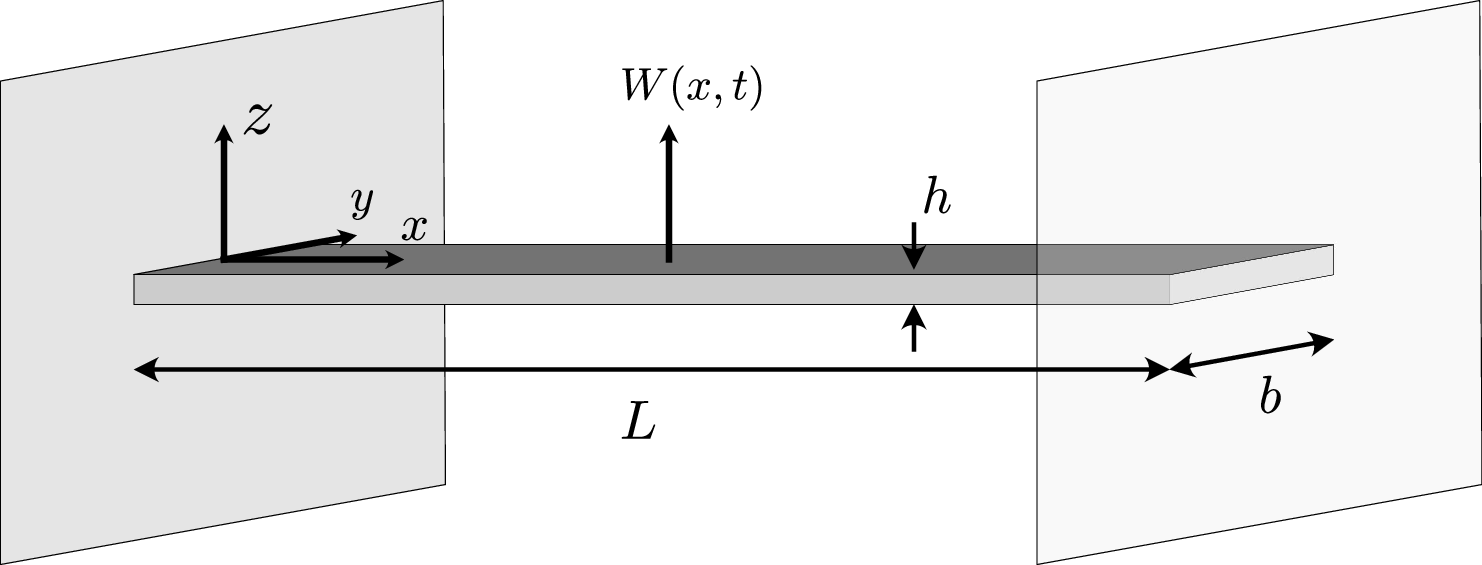}
\end{center}
\caption{A long and slender beam of length $L$, width $b$, and thickness $h$ with Cartesian coordinates ($x,y,z$), Young's modulus $E$, density $\rho$, and an intrinsic tension force $F_T$. The beam is attached to rigid walls at its ends and its transverse displacement at axial position $x$ at time $t$ is given by $W(x,t)$.}
\label{fig:beam}
\end{figure}

Consider a long and slender doubly-clamped beam of length $L$, width $b$, and thickness $h$ where $ L \!\gg\! b \!\gg\! h$ and without any external forces as shown in Fig.~\ref{fig:beam}. We include an intrinsic tension force that is often the result of the fabrication process that occurs when the actual length of the beam is not equal to its equilibrium length. This is described by the Euler-Bernoulli equation~\cite{landau:1959,inman:2008}
\begin{equation}
\rho A \frac{\partial^2 W(x,t)}{\partial t^2} + E I \frac{\partial^4 W(x,t)}{\partial x^4} - F_T \frac{\partial^2 W(x,t)}{\partial x^2} = 0 
\label{eq:eb1}
\end{equation}
where $W(x,t)$ is the transverse displacement of the beam at axial position $x$ at time $t$. In addition, $E$ is the Young's modulus, $I \!=\! b h^3/12$ is the area moment of inertia, $A \!=\! b h$ is the cross-sectional area, $\rho$ is the beam density, and $F_T$ is the axial force where the minus sign indicates tension. The boundary conditions for a doubly-clamped beam are zero displacement at the boundaries with the beam approaching the boundaries horizontally. This can be expressed as $W(0,t) \!=\! W(L,t) \!=\! W'(0,t) \!=\! W'(L,t) \!=\!0$, where a prime indicates a derivative with respect to $x$. 

When the beam deflects or bends, this introduces an additional stretching contribution to the tension term $\Delta F_T$. For small deflections, this can be represented as $\Delta F_T \!=\! \frac{\Delta L}{L} E A$ where $\Delta L$ is the added length of the beam due to its deformation and $\Delta L/L$ is the strain. This stretching of the beam, due to $W(x,t)$, can be expressed using geometric arguments~\cite{lifshitz:2008} as 
\begin{equation}
\Delta F_T = \frac{E A}{2 L} \int_0^L \left( \frac{\partial W}{\partial x} \right)^2  dx.
\end{equation}
Writing Eq.~(\ref{eq:eb1}) in terms of the intrinsic and stretching induced tension contributions, and introducing the nondimensional $x$ coordinate $x^* \!=\! x/L$, yields
\begin{equation}
\rho A \frac{\partial^2 W(x^*,t)}{\partial t^2} + \frac{E I}{L^4} \frac{\partial^4 W(x^*,t)}{\partial x^{*4}} - \frac{1}{L^2} \left[ F_T + \frac{E A}{2 L^2} \int_0^1 \left( \frac{\partial W}{\partial x^*} \right)^2  dx^* \right] \frac{\partial^2 W(x^*,t)}{\partial x^{*2}} = 0.
\label{eq:beam3}
\end{equation}
The term in brackets is the tension contribution with the intrinsic tension followed by the stretching induced tension. The stretching induced tension is the source of the nonlinearity. In the following, we will assume that $x$ represents the nondimensional coordinate, and we will omit the $*$ notation to keep the notation simple. 

The orthogonal mode shapes of the beam, $\phi_n(x)$, where $n$ is the mode number, are determined by solving the linear problem given by Eq.~(\ref{eq:eb1}) which includes the intrinsic tension but not the stretching induced tension. The specific form of $\phi_n(x)$ depends upon the boundary conditions and the relative importance of tension.  For now, we leave $\phi_n(x)$ arbitrary in order for the expressions to remain general. $\phi_n(x)$ will be specified later when discussing specific situations of direct interest. We will assume that the $\phi_n(x)$ have been normalized such that their magnitude is unity at the antinode closest to the axial midpoint of the beam~\cite{lifshitz:2008} where $x\!=\!1/2$.

\subsection{Weakly nonlinear analysis and the Duffing oscillator}

The beam displacement can be expressed as the eigenexpansion $W(x,t) \!=\!\! \sum_{n=1}^\infty w_n(x,t)$, where the spatial and temporal dependencies of the eigenmodes can be separated as $w_n(x,t) \!=\! z_n(t) \phi_n(x)$. We highlight that $z_n(t)$ is precisely the amplitude of mode $n$ of the beam when measured at an antinode near its center ($x\!=\!1/2$) due to our choice of normalization.

We now assume that the beam is oscillating at one of its eigenmodes $w_n$. Inserting $W(x,t)\!=\!w_n(x,t)$ into Eq.~(\ref{eq:beam3}), multiplying by $\phi_n(x)$, and integrating over the beam length yields~\cite{lifshitz:2008,postma:2005} 
\begin{equation}
\ddot{z}_n(t) + \left[ \frac{E I}{\rho A L^4} \frac{\int_0^1 {\phi''}^2_n(x) dx}{\int_0^1 \phi^2_n(x) dx} 
+ \frac{F_T}{\rho A L^2} \frac{\int_0^1 {\phi'}^2_n(x) dx}{\int_0^1 \phi^2_n(x) dx } \right] z_n(t) + \left[ \frac{E}{2 \rho L^4} \frac{\left(\int_0^1 {\phi'}^2_n(x) dx \right)^2}{\int_0^1 \phi^2_n(x) dx} \right] z^3_n(t) = 0
\label{eq:beam10}
\end{equation}
where a dot is a time derivative and a prime is a spatial derivative. This is a Duffing oscillator, $\ddot{z}_n(t) + \omega_n^2 z_n(t) + \alpha_n z^3_n(t) = 0$, where it will be useful to express the natural frequency as $\omega_n^2 \!=\! \omega_{n,\beta}^2 \!+\! \omega_{n,\tau}^2$.
The bending contribution to the frequency is
\begin{equation}
\omega_{n,\beta}^2 = \frac{E I}{\rho A L^4} \frac{\int_0^1 {\phi''}^2_n(x) dx}{\int_0^1 \phi^2_n(x) dx} 
\label{eq:omegan-bending}
\end{equation}
and the intrinsic tension contribution to the frequency is 
\begin{equation}
\omega_{n,\tau}^2 =  \frac{F_T}{\rho A L^2} \frac{\int_0^1 {\phi'}^2_n(x) dx}{\int_0^1 \phi^2_n(x) dx}.
\label{eq:omegan-tension}
\end{equation}
The strength of the Duffing nonlinearity, $\alpha_n$, is 
\begin{equation}
\alpha_n = \frac{E}{2 \rho L^4} \frac{\left[\int_0^1 {\phi'}^2_n(x) dx \right]^2}{\int_0^1 \phi^2_n(x) dx}
\label{eq:alphan}
\end{equation}
which is entirely due to stretching induced tension and can be represented nondimensionally as $\bar{\alpha}_n = \left( \frac{E}{2 \rho L^4} \right)^{-1} \alpha_n$. Once $\phi_n(x)$ is known, $\omega_n$ and $\alpha_n$ can be determined using Eqs.~(\ref{eq:omegan-bending})-(\ref{eq:alphan}).

We are interested in the dynamics of a micro or nanoscale beam when driven strong enough that the amplitude of oscillation is sufficient to drive the beam into the nonlinear regime. This can be studied using a weakly nonlinear analysis on the Duffing oscillator with an external drive and a phenomenological damping term~\cite{nayfeh:1979,postma:2005,kozinsky:2006} which can be expressed as
\begin{equation}
\ddot{z}_n(t) + \frac{\omega_n}{Q_n} \epsilon \dot{z}_n(t) + \omega_n^2 z(t) + \epsilon \alpha_n z_n^3(t) = \epsilon F_d \cos (\omega_d t).
\label{eq:duffing}
\end{equation}
The damping, Duffing nonlinearity, and driving terms are included perturbatively where $\epsilon$ is a small parameter. The damping term is expressed using the quality factor $Q_n$ of mode $n$ where the quality represents the ratio of energy stored to the energy dissipated by the oscillator for a single oscillation. The quality factor can be expressed as $Q_n \!=\! m_n \omega_n / \gamma_n$ where $m_n$ is the effective mass of mode $n$ and $\gamma_n$ is the damping coefficient for mode $n$. $F_d$ is the magnitude of the driving force per unit mass and $\omega_d$ is the driving frequency.

A multiple time-scale analysis~\cite{nayfeh:1979,postma:2005,kozinsky:2006} of Eq.~(\ref{eq:duffing}) yields the following relation for steady state motion
\begin{equation}
\left[  \frac{\omega_n^2}{4 Q_n^2} + \left( \sigma_n - \frac{3}{8} \frac{\alpha_n}{\omega_n} a_n^2 \right)^2 \right] a_n^2 =  \frac{F_d}{4 \omega_n^2}
\label{eq:steady-state-motion}
\end{equation}
where $a_n$ is the steady state amplitude of mode $n$ and $\sigma_n$ is the detuning frequency of the drive of mode $n$. The detuning, $\omega_d \!=\! \omega_n \!+\! \epsilon \sigma_n$,  measures the difference between the driving frequency and the resonance of mode $n$. Equation~(\ref{eq:steady-state-motion}) can be solved for the variation of the detuning frequency, $\sigma_{p,n}$,  when measured at the peak amplitude of mode $n$, $a_{p,n}$, to yield $\sigma_{p,n} \!=\! \frac{3}{8} \frac{\alpha_n}{\omega_n} a_{p,n}^2$.

The critical amplitude of oscillation, $a_{c,n}$, occurs when the driving is sufficient to yield a single point on the amplitude-detuning curve where $\frac{d a_{n}}{d \sigma_n}$ becomes infinite~\cite{postma:2005} to yield $a_{c,n} \!=\! \frac{2 \sqrt{2}}{3^{3/4}} \frac{\omega_n}{\sqrt{Q_n \alpha_n}}$. It is useful to clearly draw attention to the different physical phenomena contained in $a_{c,n}$. The intrinsic tension and bending contributions are captured by $\omega_n$ and the stretching induced tension is included by $\alpha_n$. This can be simplified further using the expressions for $\omega_n$ and $\alpha_n$ to yield
\begin{equation}
a_{c,n} = \frac{2}{3^{5/4}} \frac{h}{\sqrt{Q_n}} \frac{\left( \int_0^1 {\phi''}^2_n(x) dx + 2 U  \int_0^1 {\phi'}^2_n(x) dx \right)^{1/2}}{\int_0^1 {\phi'}^2_n(x) dx}
\label{eq:acn2}
\end{equation}
where $U$ is the nondimensional tension parameter, $U \!=\! \frac{F_T}{2 E I/L^2}$, which represents the ratio of the intrinsic tension force to an elastic force scale. It is insightful to highlight the different contributions in Eq.~(\ref{eq:acn2}). The first term in the numerator is due to bending, the second term in the numerator is due to intrinsic tension, and the denominator is due to stretching induced tension.

The critical peak amplitude occurs when $a_{p,n}$ is 0.745 of the critical amplitude $a_{c,n}$ (this value of $a_{p,n}$ corresponds to the 1 dB compression point)~\cite{postma:2005}. We denote the root-mean-squared (RMS) value of the amplitude $a_{p,n}^*$ as $a_{p,n}^* \!=\! \frac{0.745}{\sqrt{2}} \, a_{c,n}$
where the $\sqrt{2}$ is from the conversion of $a_{c,n}$ from an amplitude to its RMS value. In an experiment, the beam is in the linear dynamic regime when it is driven such that $a_{th,n} \!\le\! a_{p,n} \!\le\! a_{p,n}^*$ where $a_{th,n}$ is the RMS amplitude of vibration due to thermal motion alone. This choice implies that the experimental measurement bandwidth is much larger than the natural bandwidth of the mode.

Equation~(\ref{eq:acn2}) is a very general and useful result. In many cases of interest,  $\phi_n(x)$ and the properties $h$ and $U$ are known. However, $Q_n$ depends upon the details of the experiment and, as a result, there is not a general expression that can be used. Most importantly, $Q_n$ depends on the effective damping or dissipation acting on the mode which can have subtle contributions from a variety of phenomena. There are intrinsic mechanisms of dissipation due to bending and friction and extrinsic mechanisms of dissipation due to acoustic radiation from the beam into the supports~\cite{cross:2001:prb}, soft-clamping losses~\cite{unterreithmeier:2010}, the presence of surface defects~\cite{villanueva:2014}, and the interactions with a surrounding fluid if present~\cite{sader:1998,paul:2004,paul:2006}.

When the beam is placed in a vacuum, the effective mass of each mode can be found by ensuring that the kinetic, or potential, energy of the oscillating mode equals that of the oscillating beam with that mode shape~\cite{ari:2021,gress:2023}. Nanostructures composed of silicon nitride, that are under significant intrinsic tension, are well known for their extremely high quality factors~\cite{engelsen:2024}, where values achieving $Q \!>\! 10^9$ have been realized~\cite{bereyhi:2022}. These extreme quality factors are due to damping dilution; there is a rich literature investigating its physical origins and application to new technologies~\cite{bachtold:2022,engelsen:2024}.

When the oscillator is immersed in a fluid, such as air or water, the effective mass and the effective damping acting on the modes are frequency dependent~\cite{sader:1998,paul:2006}. As a result, our analysis presented here, which assumes that both the effective mass and damping are constant for each mode, would not directly apply. However, it is interesting to highlight that when a fluid is present, the fluid dissipation typically far exceeds the intrinsic and extrinsic damping mechanisms described above. Furthermore, the quality factor of a beam immersed in a fluid \emph{increases} with mode number $n$ while also lowering the resonant frequencies from their natural values~\cite{villa:2009,clark:2010}.

Therefore, if $Q_n$ is known, which can often be measured experimentally or estimated for a particular situation of interest, then we have the desired result for $a_{c,n}$.  An expression of more general use is
\begin{equation}
\frac{a_{c,n}}{h} \sqrt{Q_n} = \frac{2}{3^{5/4}} \frac{\left[ \int_0^1 {\phi''}^2_n(x) dx + 2 U  \int_0^1 {\phi'}^2_n(x) dx \right]^{1/2}}{\int_0^1 {\phi'}^2_n(x) dx}
\label{eq:acn-Qn}
\end{equation}
where the right hand side is completely determined by $\phi_n(x)$ and $U$. Equation~(\ref{eq:acn-Qn}) is a nondimensional scaled amplitude that is independent of the quality factor. We will use Eq.~(\ref{eq:acn-Qn}) in our study of the role of tension, bending, and the boundary conditions on the critical amplitude.

\subsection{The thermal fluctuations of oscillator displacement}

The lower bound of the linear dynamic range is set by the magnitude of the displacement fluctuations due to thermal motion. The amplitude of deflection for mode $n$ due to thermal fluctuations, $a_{th,n}$, can be found using the equipartition of energy~\cite{sethna:2006}, $\frac{1}{2} k_B T \!=\! \frac{1}{2} k_n \langle z_n^2 \rangle$, where each quadratic mode has an energy of $k_B T/2$. $z_n(t)$ is the stochastic fluctuations of mode $n$ of the beam, $k_n$ is the mode-dependent spring constant, $k_B$ is Boltzmann's constant, $T$ is the temperature, and the angle brackets indicate an equilibrium ensemble average. This can be rearranged to yield $a_{th,n} \!=\! \sqrt{\frac{k_B T}{k_n}}$ 
where $a_{th,n} \!=\! \langle z_n^2 \rangle ^{1/2}$ is the RMS of the thermal vibrations. Since it is expected that the higher modes will be stiffer with increasing $n$, the amplitude of thermal motion decreases with increasing mode number.

The linear dynamic range of mode $n$ is defined as~\cite{postma:2005}
\begin{equation}
\mathcal{L}_n = \frac{a^*_{p,n}}{a_{th,n}}.
\label{eq:ldr}
\end{equation}
For an oscillator with $\mathcal{L}_n \!=\! 1$, the critical peak amplitude of mode $n$   equals the amplitude of the thermal fluctuations. This oscillator would not have any region of parameters for which the dynamics would be linear. The interesting limit where $\mathcal{L}_n \!\le\! 1$ has been probed using carbon nanotubes~\cite{barnard:2019} and a nanoparticle placed in an optical trap~\cite{gieseler:2013}.  For $\mathcal{L}_n \!>\! 1$ there are a range parameters where the amplitude of the driven oscillator is larger than the thermal fluctuations and smaller than the critical amplitude upon which nonlinearity is important. This oscillator would have a regime of operating conditions with linear dynamics. It will be important to understand the variation of $\mathcal{L}_n$ and $a_{th,n}$ with parameters for the clamped beam, string, and hinged beam.

The prediction of the magnitude of the thermal fluctuations requires $k_n$. The modal spring constant can be expressed as $k_n \!=\! m_n \omega_n^2$ where $m_n$ is the effective mass of mode $n$.  The effective mass can be determined by first expressing the motion of the mode $n$ at frequency $\omega_n$ as $W_n(x,t) = c_n \phi_n(x) e^{i \omega_n t}$ where $c_n$ is constant. Equating the kinetic energy of the lumped mass and the spatially extended beam yields  
\begin{equation}
\frac{1}{2} m_n \dot{W}_n(x_0,t)^2 = \frac{1}{2} L \int_0^1 \rho A \dot{W}_n(x,t)^2 dx.
\label{eq:ke}
\end{equation}
Using the expression for $W_n(x,t)$ and rearranging yields $ m_n \!=\! m \int_0^1  \phi_n(x)^2 dx$ where $m$ is the total mass of the beam, $m\!=\!\rho L b h$, and $\phi_n(x_0)=1$ using our normalization since we assume the measurement location $x_0$ is at the antinode near the center of the beam.  The spring constant is then $k_n \!=\! m \omega_n^2 \int_0^1 \phi_n(x)^2 dx$. We emphasize that the intrinsic tension is included through its influence on the mode shape $\phi_n(x)$.
The amplitude of thermal motion can now be expressed as
\begin{equation}
a_{th,n} = \sqrt{\frac{k_B T}{m \omega_n^2}} \left(\int_0^1  \phi_n(x)^2 dx \right)^{-1/2}.
\label{eq:athn}
\end{equation}
The linear dynamic range $\mathcal{L}_n$ can now be expressed using the expressions for $a_{c,n}$  and $a_{th,n}$ in Eq.~(\ref{eq:ldr}).

\section{Exploring the linear dynamic range}
\label{section:ldr}

\subsection{A beam with clamped boundaries}
\label{section:clamped}

We start by quantifying the linear dynamic range of an elastic beam under intrinsic tension with clamped boundary conditions. The mode shapes are determined by solving Eq.~(\ref{eq:eb1}) with clamped boundary conditions to yield~\cite{bokaian:1990,stachiv:2014,ari:2021}
\begin{equation}
\phi_n(x) = \frac{1}{N_n^*} \left[ \cosh \left(M_n x\right) - \cos \left(N_n x\right) - C_0 \left( \sinh \left(M_n x \right) - \frac{M_n}{N_n} \sin \left(N_n x \right) \right) \right]
\label{eq:phi-clamped-clamped}
\end{equation}
where $N_n^*$ is the mode dependent normalization constant such that the mode shape's maximum value is unity at the antinode closest to the middle of the beam ($x\!=\!1/2$). The dimensionless constants $M_n$ and $N_n$ are $M_n \!=\! ( U \!+\! \sqrt{U^2 \!+\! \Omega_n^2} )^{1/2}$ and $N_n \!=\! ( -U \!+\! \sqrt{U^2 \!+\! \Omega_n^2} )^{1/2}$.   The constant $C_0$ is given by $C_0 \!=\! \frac {\cosh(M_n) \!-\! \cos(N_n) } {\sinh (M_n) \!-\! \frac{M_n}{N_n} \sin (N_n)}$. 
The nondimensional natural frequencies,  $\Omega_n$, are the roots of the characteristic equation
\begin{multline}
\Omega_n + U \sinh\left[ \left( U + \sqrt{U^2 + \Omega_n^2} \right)^{1/2} \right] \sin\left[ \left( -U + \sqrt{U^2 + \Omega_n^2}\right)^{1/2} \right] \\ - \Omega_n \cosh \left[ \left( U + \sqrt{U^2 + \Omega_n^2}\right)^{1/2} \right] \cos \left[ \left(-U + \sqrt{U^2 + \Omega_n^2} \right)^{1/2} \right]  = 0
\label{eq:characteristic-equation}
\end{multline}
where $\Omega_n \!=\! \frac{\omega_n}{\xi/L^2}$, $\xi \!=\! \sqrt{EI/\mu}$,  and $\mu\!=\!\rho b h$ is the mass per unit length. It is insightful to highlight that $C_0 \approx 1$ for $U \!\gtrsim\! 10$.  As a result of the complexity of the expression for $\phi_n(x)$ given by Eq.~(\ref{eq:phi-clamped-clamped}), the expressions for $\omega_n$ and $\alpha_n$ need to be evaluated numerically for a given value of $U$ in order to determine $a_{c,n}$.

In Eqs.~(\ref{eq:phi-clamped-clamped})-(\ref{eq:characteristic-equation}), $U$ represents the ratio of the intrinsic tension force to an elastic or bending force.  When $U \!=\! 0$, this yields the classic Euler-Bernoulli beam result, and when $U \!\gg\! 1$ the results approach that of a string. We will discuss in more detail the high tension limit and its connection with the string for the higher modes in \S\ref{section:string}.

\begin{figure}[h!]
\vspace{1cm}
\begin{center}
\includegraphics[width=3in]{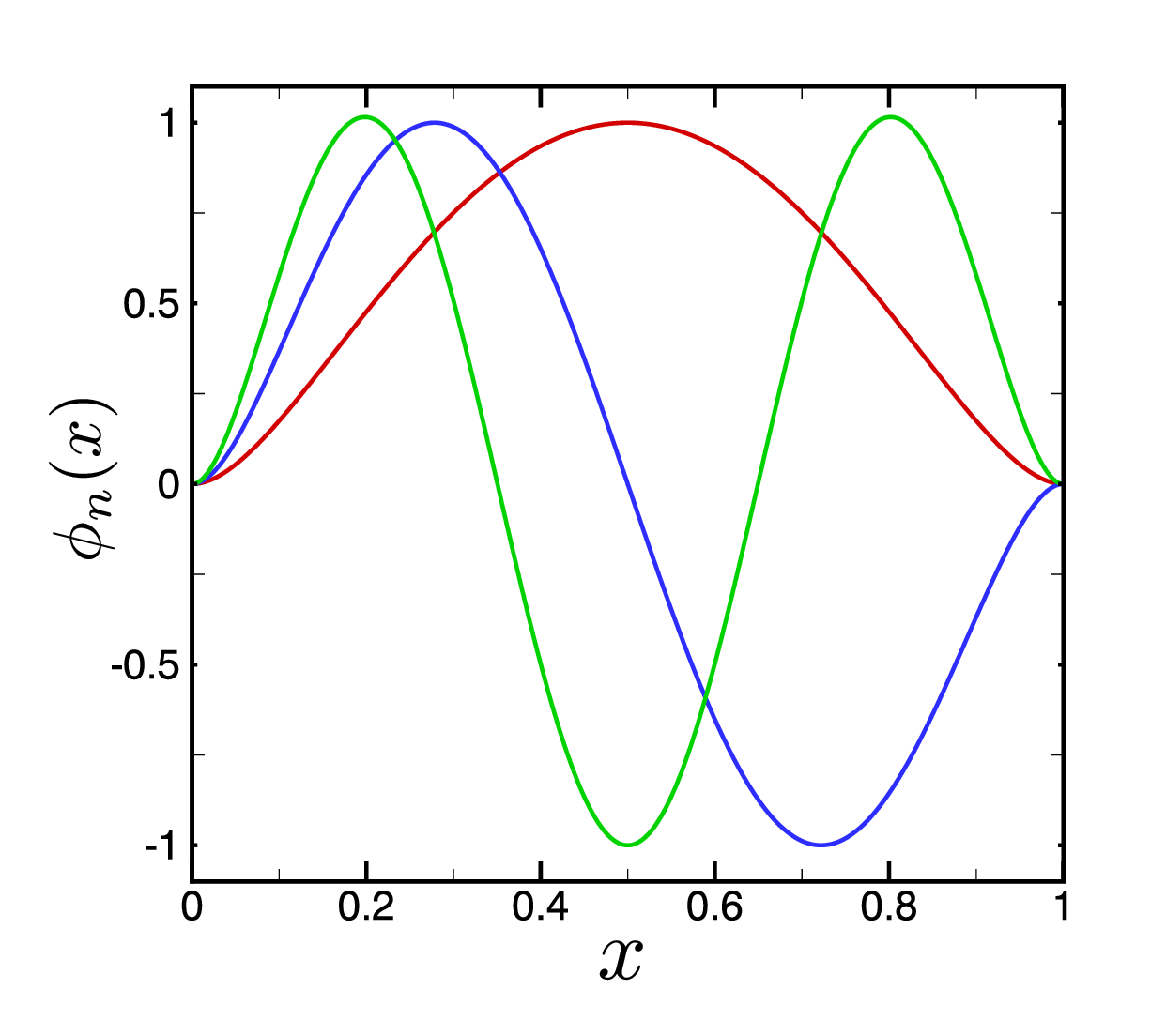}
\end{center}
\caption{The first three mode shapes $\phi_n(x)$ for a doubly-clamped beam with a tension parameter of $U\!=\!100$ where $n\!=\!1$ (red), $n\!=\!2$ (blue), $n\!=\!1$ (green).}
\label{fig:phin}
\end{figure}

The first three mode shapes of a beam given by Eq.~(\ref{eq:phi-clamped-clamped}) with $U\!=\!100$, are shown in Fig.~\ref{fig:phin}. The clamped boundary condition requires that the beam remain horizontal at the two ends which causes the mode shape to have a sharp structure near the boundaries where $|\frac{d \phi_n}{dx}|$ becomes large with a magnitude that increases with increasing mode number and with increasing tension. 

\begin{figure}[h!]
\vspace{1cm}
\begin{center}
\includegraphics[width=3.2in]{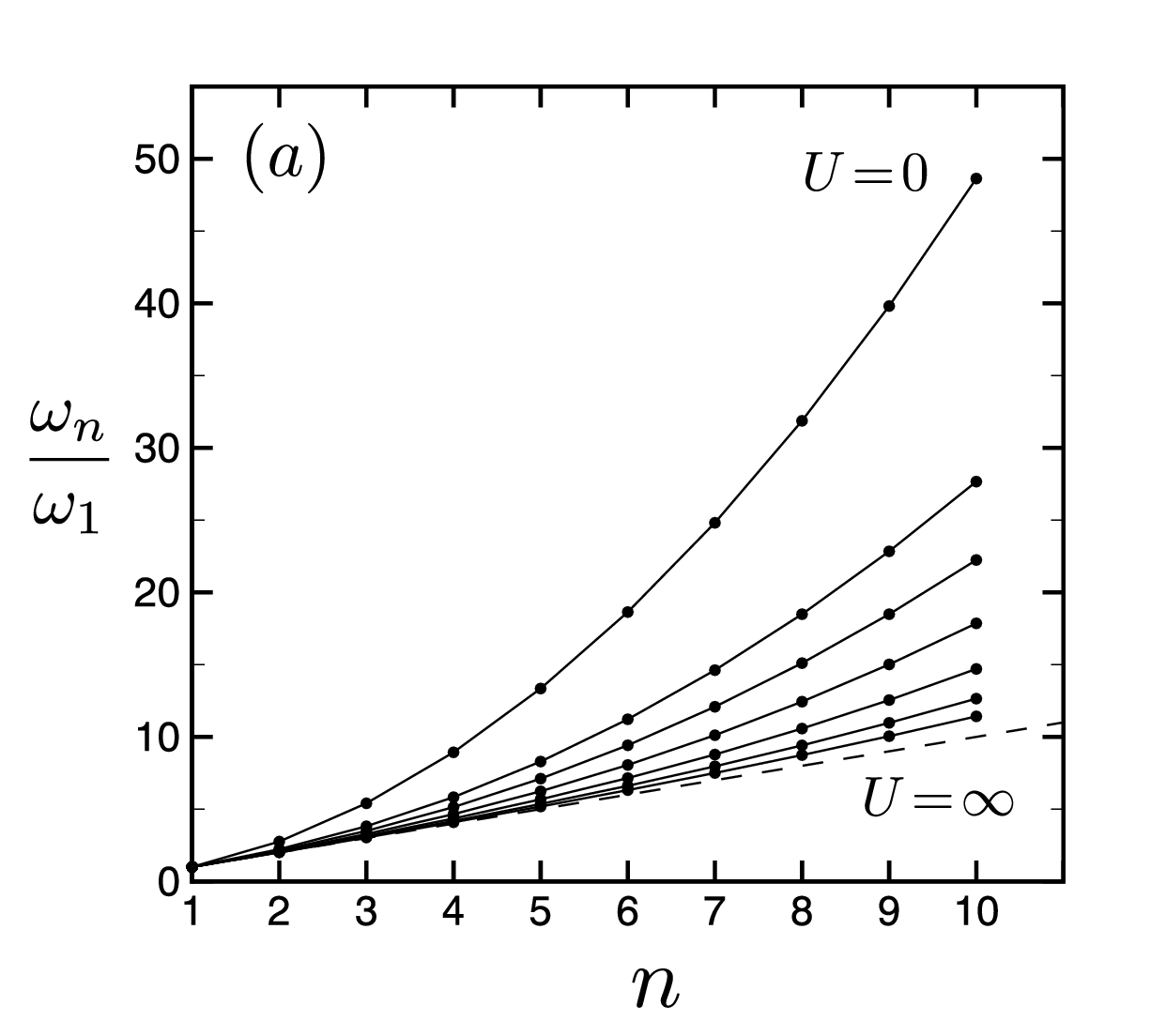}
\includegraphics[width=3.2in]{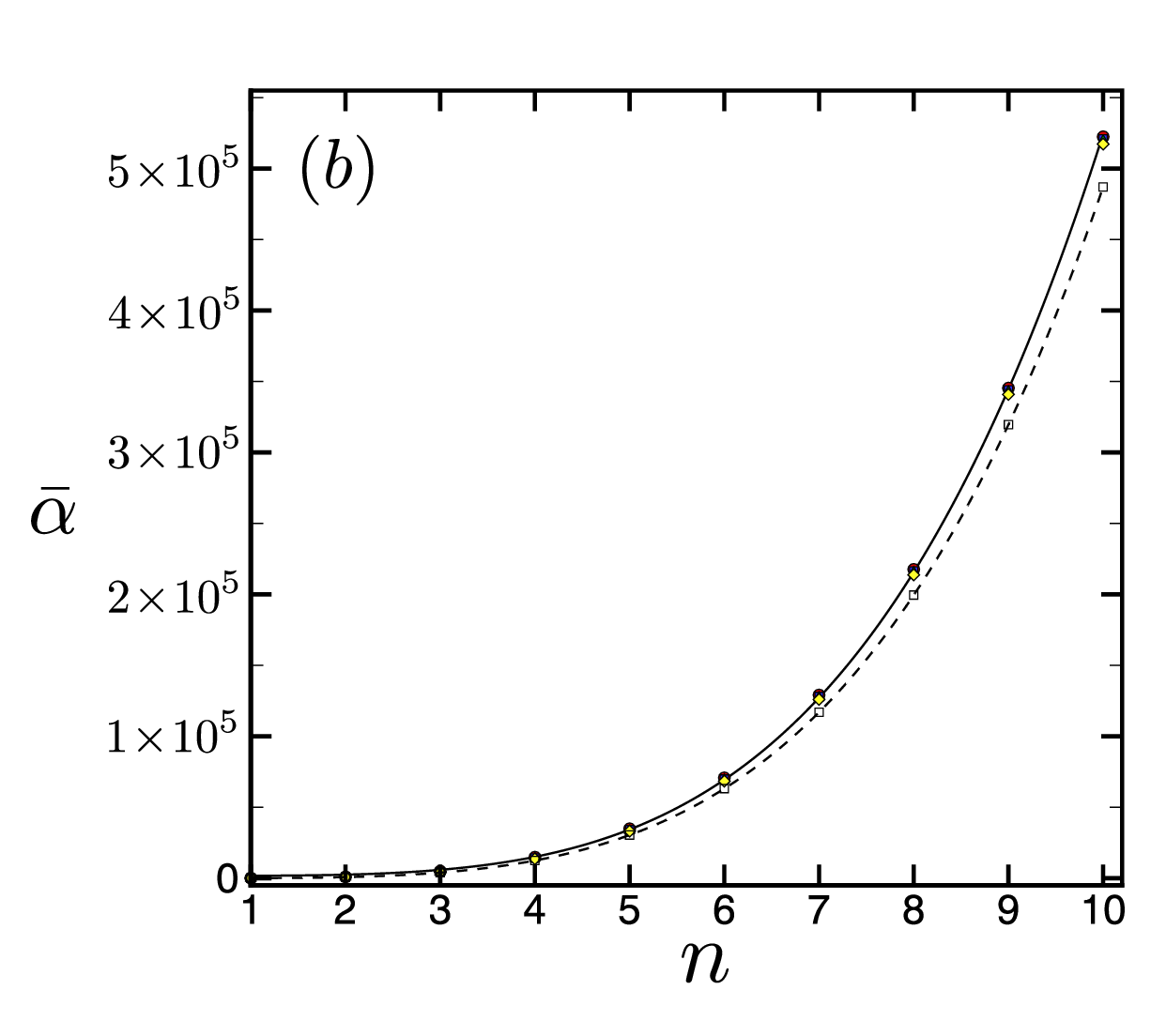}
\end{center}
\caption{Natural frequency and Duffing parameter for a doubly-clamped beam.  ($a$)~Variation of  $\omega_n/\omega_1$ with $n$ and $U$. Top curve ($U\!=\!0$) and the following solid curves are  $U\!=\!50,100,200,400,800,1600$ in sequential order. Dashed line is asymptotic result for $U\!=\!\infty$ where $\omega_n/\omega_1\!=\!n$. ($b$)~Variation of the nondimensional Duffing parameter $\bar{\alpha}$ with $n$ for: $U\!=\!0$ (red, circles), $U\!=\!50$ (blue, squares), $U\!=\!100$ (green, triangles), and $U\!=\!200$ (yellow, diamonds). The variation of $\bar{\alpha}_n$ is insensitive to $U$ as indicated by the collapse of the results onto a single curve. The solid line is $\bar{\alpha} \!=\! 1.67 \!\times10^{3}\! + \! 52.4 \, n^4$ yielding  $\bar{\alpha} \!\propto\! n^4$. The dashed line is for a string where $\bar{\alpha}_n \!=\! \frac{\pi^4}{2} n^4$.}
\label{fig:omegan_over_omega1}
\end{figure}

The natural frequencies, $\omega_n$, can be determined either by directly solving Eqs.~(\ref{eq:omegan-bending})-(\ref{eq:omegan-tension}) with the mode shape from Eq.~(\ref{eq:phi-clamped-clamped}) or by solving Eq.~(\ref{eq:characteristic-equation}) for $\Omega_n$ and using the definition of $\Omega_n$. The variation of $\omega_n/\omega_1$ with the tension parameter $U$ is shown in Fig.~\ref{fig:omegan_over_omega1}(a) for the first ten modes. As $U$ increases, the variation of $\omega_n/\omega_1$ approaches the linear result $\omega_n/\omega_1\!=\!n$ which describes a string.

The variation of the nondimensional Duffing parameter $\bar{\alpha}$ is shown in Fig.~\ref{fig:omegan_over_omega1}(b). Results for several values of $U$ are shown, $0 \!\le\! U \!\le\! 200$, using different symbols. The results collapse to a single curve indicating that $\bar{\alpha}$ is independent of $U$. The solid line is a curve fit through the data which yields the dependence $\bar{\alpha} \propto n^4$.

\begin{figure}[h!]
\vspace{1cm}
\begin{center}
\includegraphics[width=3.5in]{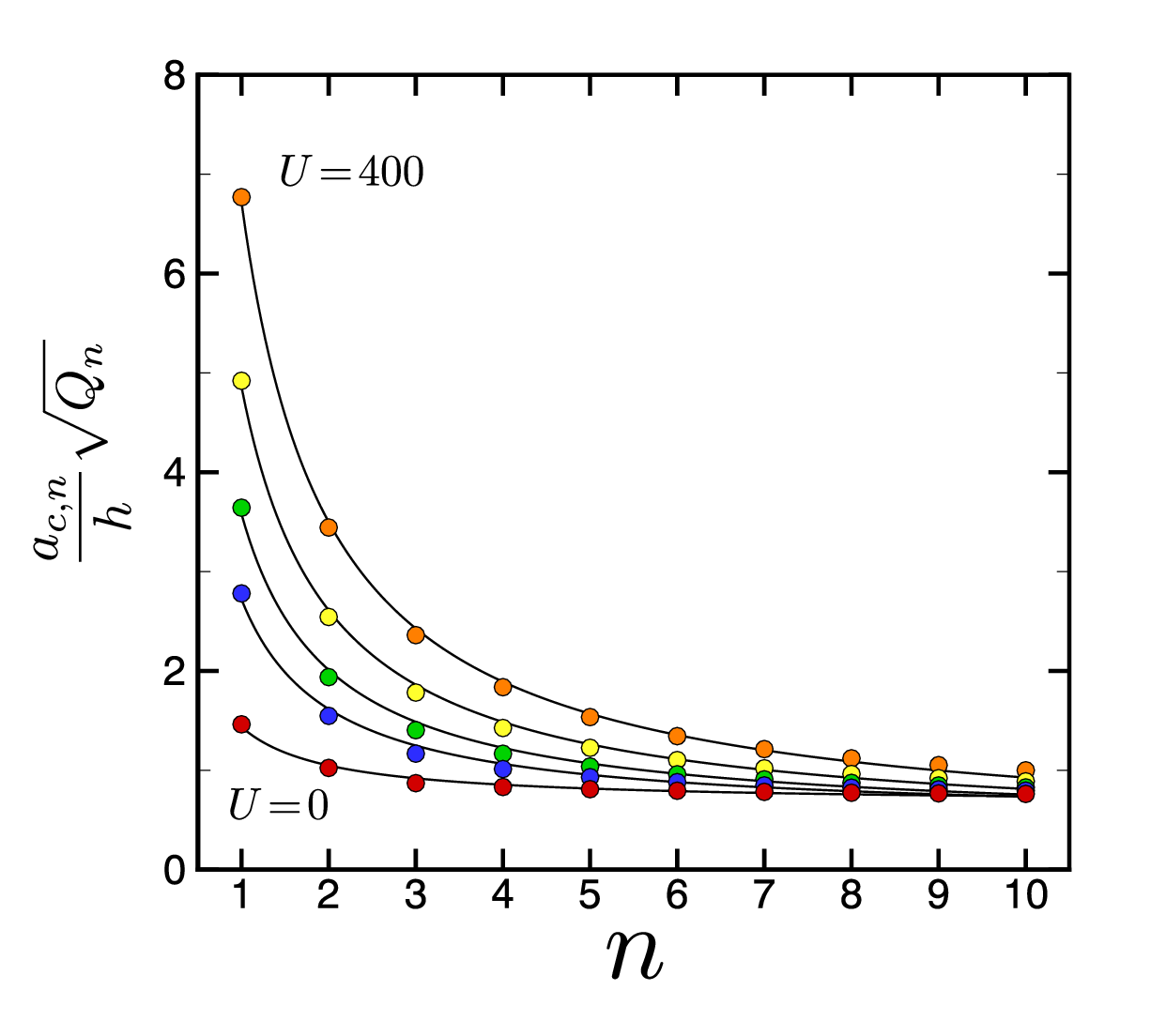}
\end{center}
\caption{Variation of $a_{c,n} \frac{\sqrt{Q_n}}{h}$ with $n$ for a doubly-clamped beam for $U\!=\!0$ (red), $U\!=\!50$ (blue),  $U\!=\!100$ (green), $U\!=\!200$ (yellow), $U\!=\!400$ (orange). Solid lines are curve-fits  of the form $a_{c,n} \frac{\sqrt{Q_n}}{h} \!=\! \beta \!+\! \frac{\tau}{n}$ where $\beta$ and $\tau$ are constants.} 
\label{fig:acqoverh-clamped}
\end{figure}

In Fig.~\ref{fig:acqoverh-clamped} we show the variation of $\frac{a_{c,n}}{h} \sqrt{Q_n}$ with mode number and with the tension parameter for clamped boundaries. Results for $0 \!\le\! U \!\le\! 400$ are shown using different colors and data symbols. The solid lines are curve fits through the data of the form
\begin{equation}
\frac{a_{c,n}}{h}\sqrt{Q_n} \!=\! \beta \!+\! \frac{\tau}{n}
\label{eq:bending-tension}
\end{equation}
where $\beta$ and $U$ are constants whose values depend upon $U$.  The curve fits show excellent agreement with the theoretical predictions given by the data symbols. The values of $\beta$ and $\tau$ capture the influence of bending, intrinsic tension, and stretching induced tension.  The values of $\beta$, for all $U \!>\! 0$, approach the $U \!=\! 0$ value of $\beta \!=\! 0.739$ for $n \!\gg\! 1$ as shown in Fig.~\ref{fig:acqoverh-clamped}. To elucidate the role of tension we use a string model in \S\ref{section:string}. To clearly quantify the competing roles of tension and bending we use a hinged model in \S\ref{section:hinged}.

\begin{figure}[h!]
\vspace{1cm}
\begin{center}
\includegraphics[width=3.2in]{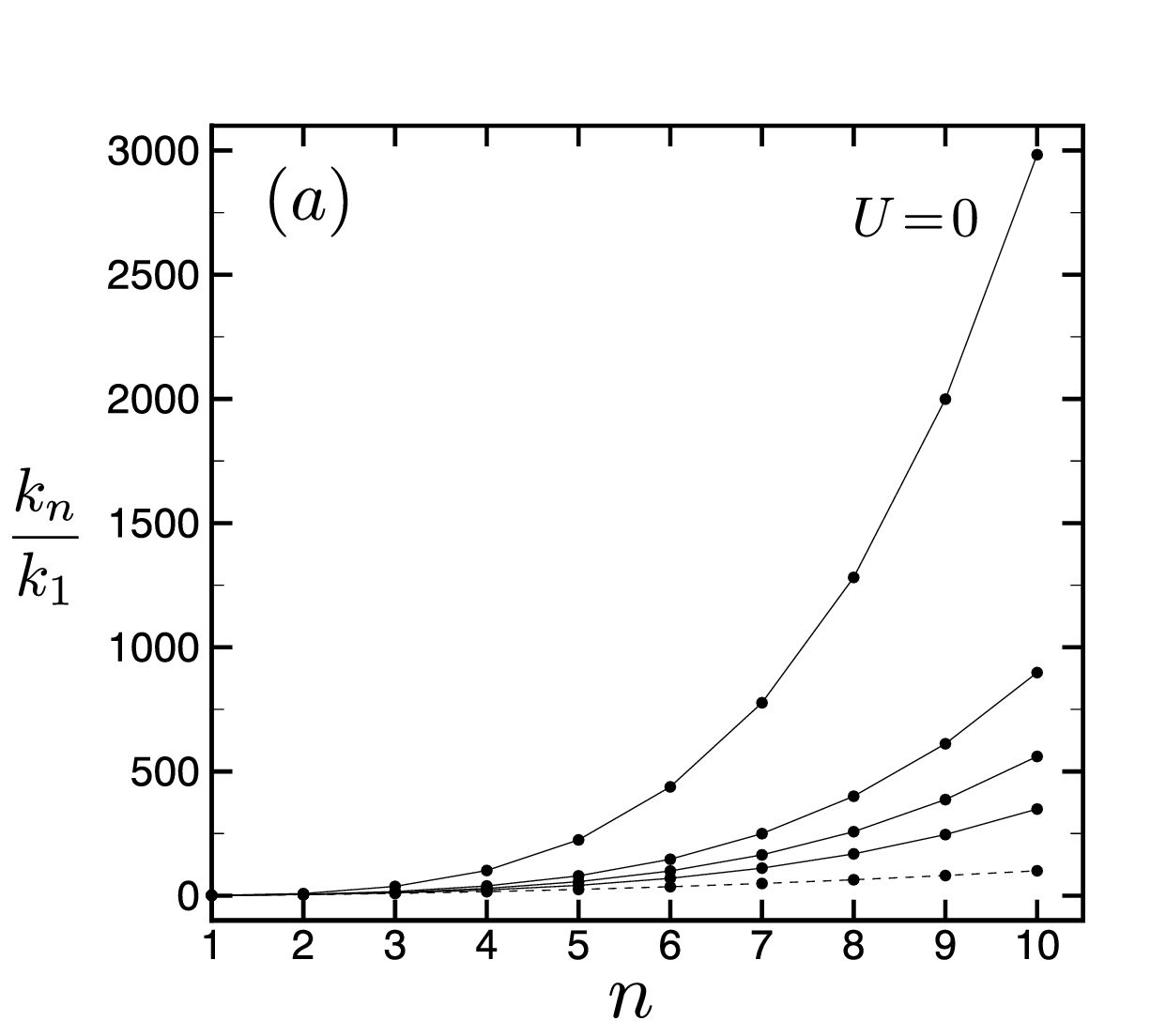}
\includegraphics[width=3.2in]{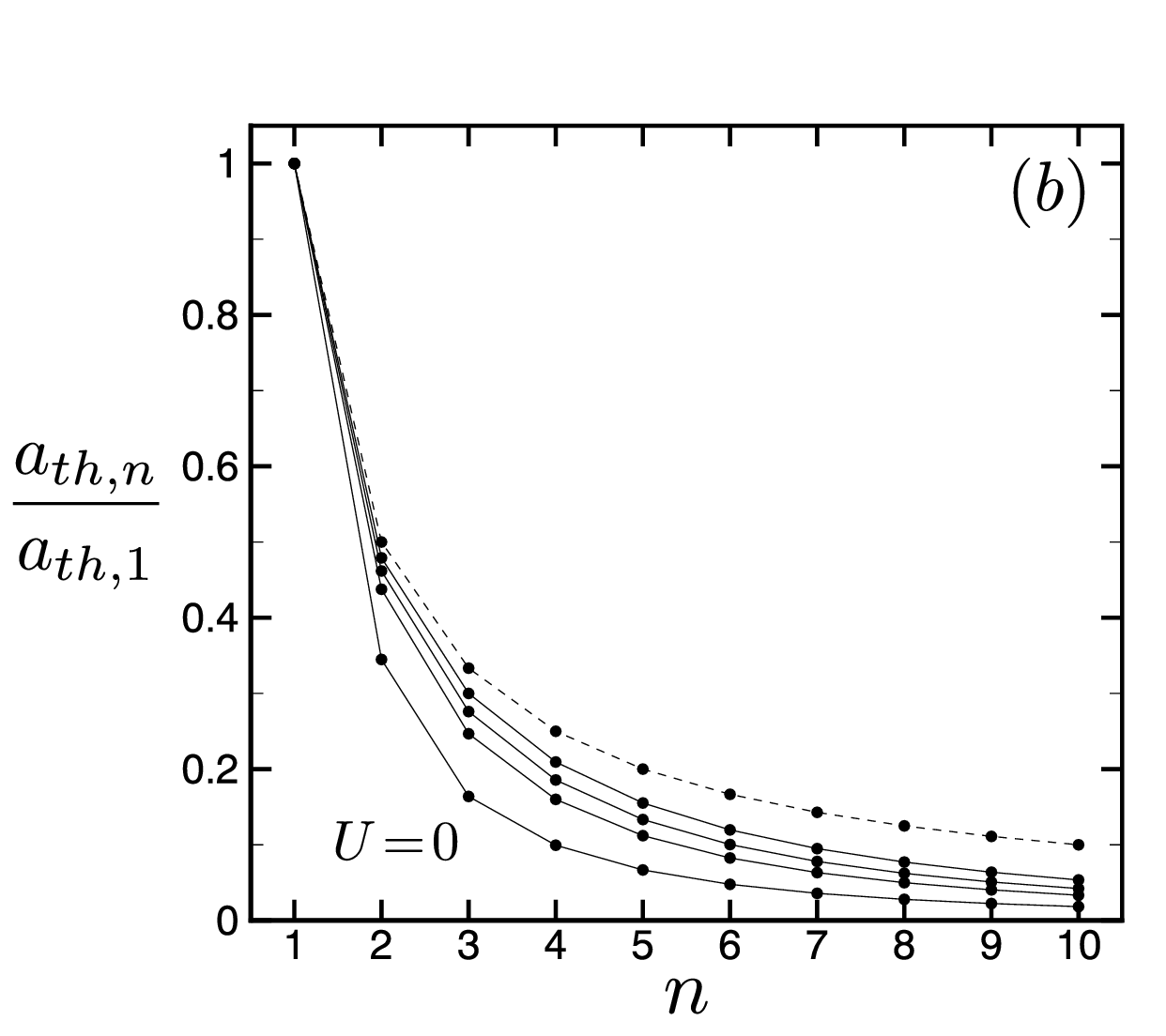}
\end{center}
\caption{($a$) Variation of the spring constant with $n$ for a beam with clamped boundaries. The top curve (labeled) is for an Euler-Bernoulli beam in the absence of intrinsic tension ($U\!=\!0$). The following curves are for $U\!=\!50,100,200$ and appear in order as the dashed line is approached. The dashed line is for $U\!=\!\infty$ where $k_n/k_1\!=\!n^2$. ($b$) Variation of the amplitude due to thermal motion with mode number and with tension. The lower curve is for $U\!=\!0$ and the remaining solid curves are for $U\!=\!50,100,200$ in order as they approach the string result shown by the dashed line. For the string, $a_{th,n}/a_{th,1} \!=\! n^{-1}$.} 
\label{fig:kn-over-k1-clamped}
\end{figure}

The modal variation of the spring constant with increasing intrinsic tension is shown in Fig.~\ref{fig:kn-over-k1-clamped}(a). The Euler-Bernoulli beam without intrinsic tension, $U\!=\!0$, is labeled. For increasing values of $U$, the magnitude of the values of $k_n/k_1$ decrease. Additional curves are shown for $U\!=\!50,100, 200$. The spring constant increases with increasing mode number and the values become more closely spaced with increasing intrinsic tension. The variation of the amplitude due to thermal motion is shown in Fig.~\ref{fig:kn-over-k1-clamped}(b). The amplitude of thermal motion decreases with increasing mode number which is due to the increasing magnitude of the spring constant shown in Fig.~\ref{fig:kn-over-k1-clamped}(a).

\subsection{The string approximation}
\label{section:string}

In the high tension limit, $U \!\gg\! 1$, if we neglect the bending contribution in favor of the tension term, the beam equation given by Eq.~(\ref{eq:eb1}) becomes
\begin{equation}
\rho A \frac{\partial^2 W(x,t)}{\partial t^2} - \frac{F_T}{L^2} \frac{\partial^2 W(x,t)}{\partial x^2} = 0 
\label{eq:string1}
\end{equation}
where we are using nondimensional $x$. Equation~(\ref{eq:string1}) describes the dynamics of a string with boundary conditions $W(0,t)\!=\!W(1,t)\!=\!0$ where there is no longer a constraint on the slope at the walls. The mode shape of the string is $\phi_n(x) = \sin \left( n \pi x\right)$ which leads to an orthogonality constant $\tilde{N}_n \!=\! 1/2$ for all $n$. 

\begin{figure}[h!]
\vspace{1cm}
\begin{center}
\includegraphics[width=3.0in]{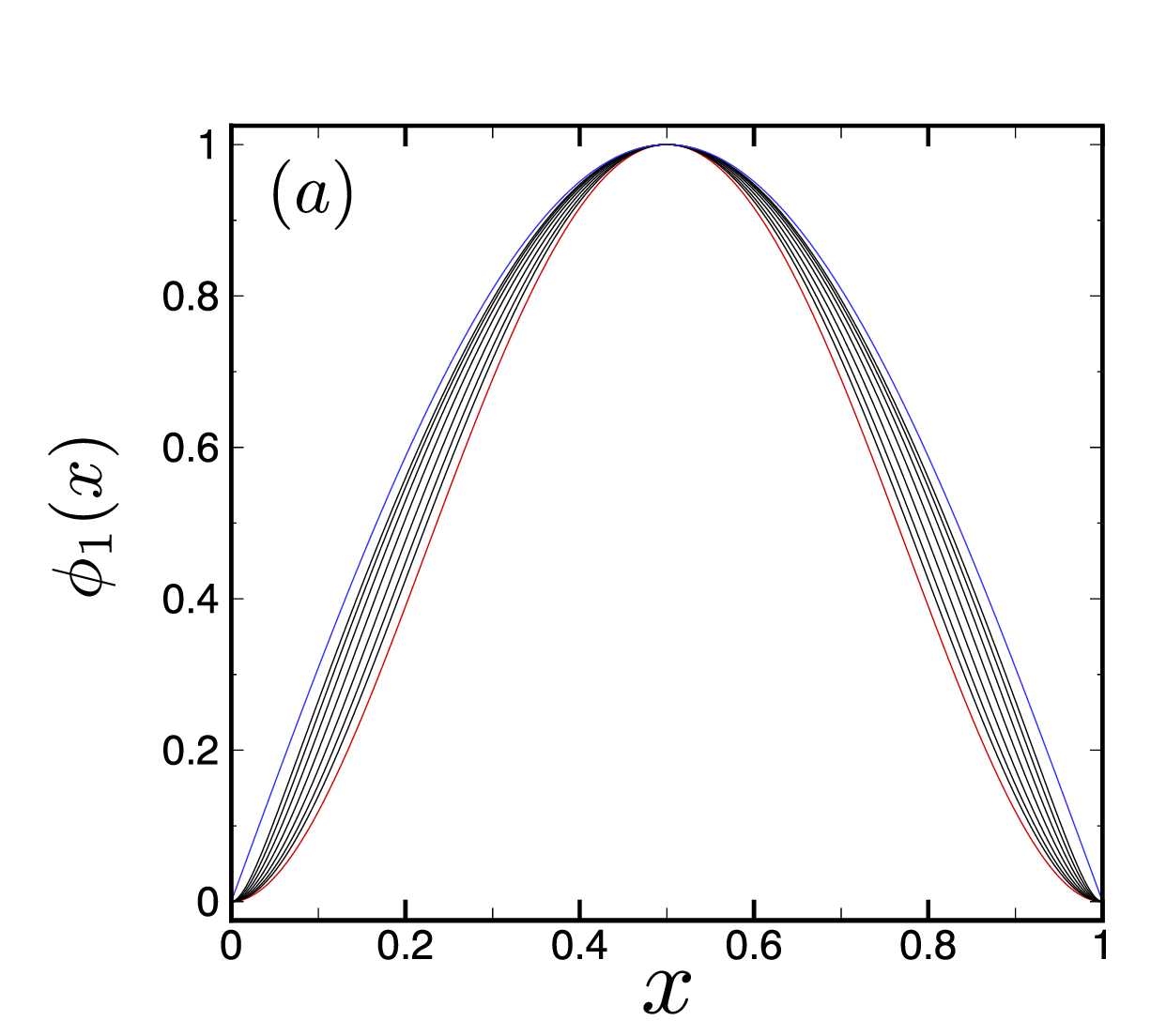}  
\includegraphics[width=3.0in]{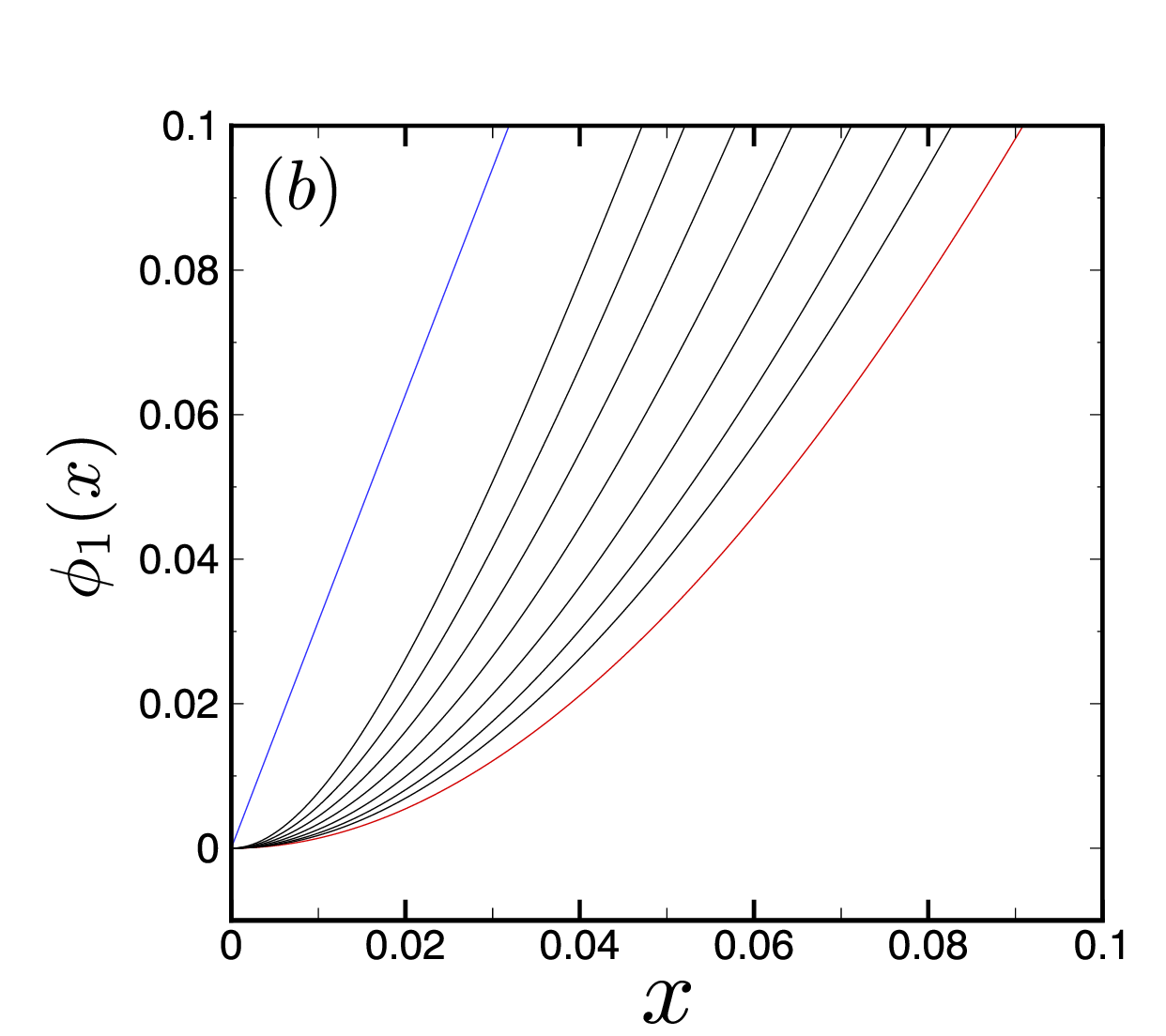} \\ 
\includegraphics[width=3.0in]{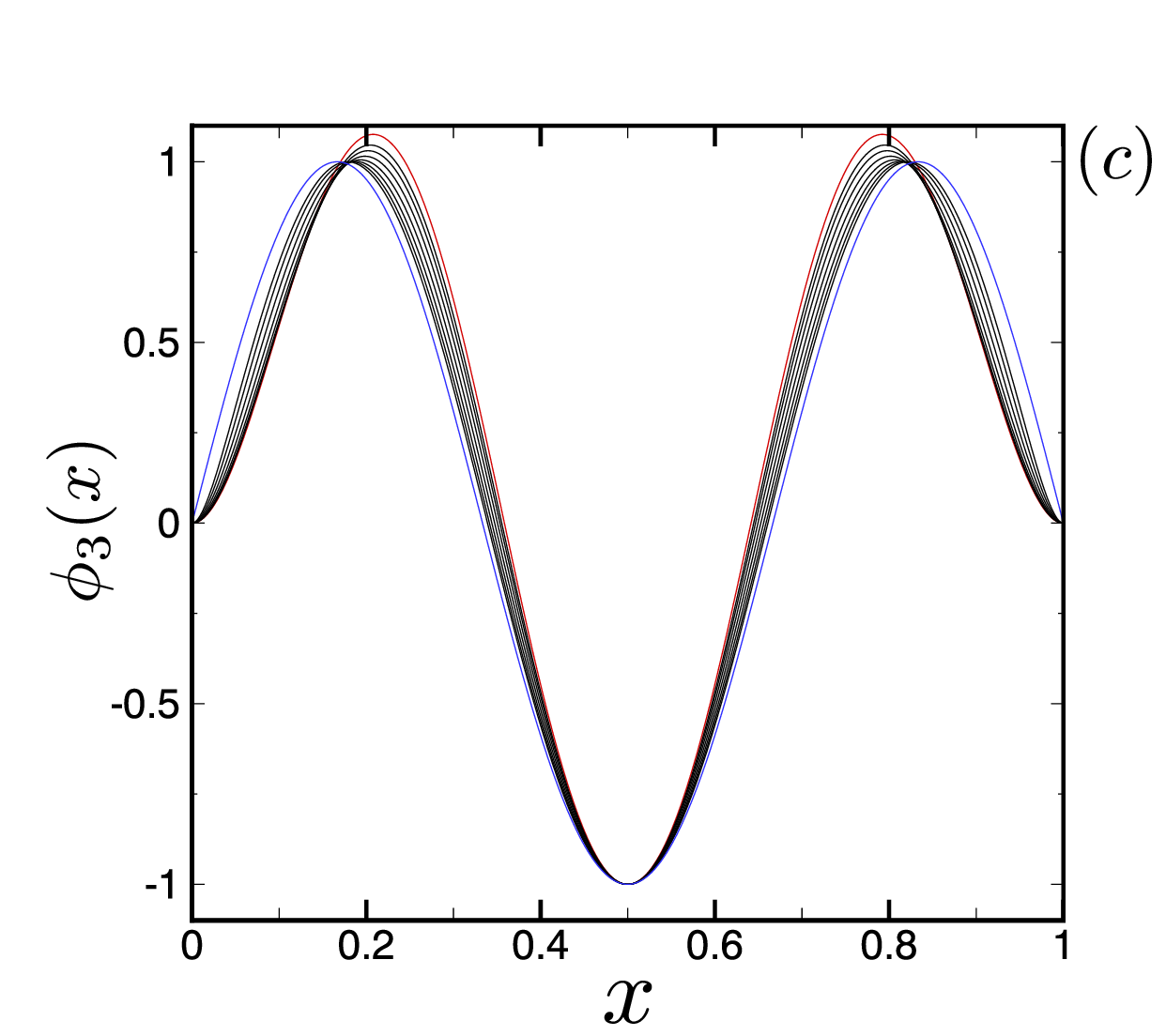}
\end{center}
\caption{(a)~Variation of $\phi_1(x)$ for a clamped beam and string. An Euler-Bernoulli beam without intrinsic tension, $U\!=\!0$, is shown in red. Black curves show $\phi_1(x)$ for a clamped beam with increasing values of $U$ where $U \!=\! 25,50, \ldots, 1600$ which occur in order when going from the red to blue curve. The blue curve is $\phi_1(x)$ for a string given which is identical to the mode shape of a beam with hinged boundaries. (b)~Close-up of the mode shapes near the boundary at $x\!=\!0$. (c)~The spatial variation of $\phi_3(x)$ using the same color conventions. The mode shapes of a hinged beam are independent of $U$ and are identical that of a string for all $n$.} 
\label{fig:phi-comp}
\end{figure}

A comparison of the string and clamped-beam mode shapes is shown in Fig.~\ref{fig:phi-comp}(a) for the fundamental mode. The fundamental mode of an Euler-Bernoulli beam without intrinsic tension is shown in red which is found using $U\!=\!0$ in Eq.~(\ref{eq:phi-clamped-clamped}). The black curves show the variation in the mode shape with increasing values of $U$ where $U \!=\! 25,50, \!\ldots\!,\! 1600$ which occur in order when going from the red curve to the blue curve representing the string mode shape. Increasing values of intrinsic tension in a clamped beam cause its  mode shape to approach that of the string while maintaining the incompatible zero slope boundary condition at the walls. A close-up view of the variation of the fundamental mode shapes near the boundary at $x\!=\!0$ is shown in Fig.~\ref{fig:phi-comp}(b). It is this observation that a clamped beam under high tension has mode shapes that are essentially sinusoidal, except very near the boundaries, which will lead to the hinged beam model we discuss in \S\ref{section:hinged}.

As an example of the mode shapes for larger $n$, we show the variation of $\phi_3(x)$ in Fig.~\ref{fig:phi-comp}(c). Using our normalization, $\phi_3(x)$ at the midpoint $x\!=\!1/2$ equals unity. It is interesting to point out that the maximum value of $\phi_n(x)$ does not always occur near $x\!=\!1/2$ for finite values of $U$ as illustrated here for the case of $n\!=\!3$.  

The sinusoidal mode shape for the string immediately yields useful analytical expressions for $\omega_n$, $\alpha_n$, and $a_{c,n}$. Using the string's mode shape in Eq.~(\ref{eq:omegan-tension}) yields 
$\omega_n \!=\! \frac{\pi n}{L} \sqrt{\frac{F_T}{\rho A}}$ where we have used $\omega_{n,\beta} \!=\! 0$. This yields $\omega_n/\omega_1=n$ for the string which is shown as the dashed line in Fig.~\ref{fig:omegan_over_omega1}(a).  The nonlinear Duffing parameter can be found using Eq.~(\ref{eq:alphan}) to yield $\alpha_n \!=\! \left(\frac{E \pi^4}{4 \rho L^4} \right) n^4$
where $\alpha_n \!\sim\! n^4$. It is important to emphasize that the Young's modulus $E$ enters the string  description here due to the stretching induced tension that causes the geometric nonlinearity.  The $n^4$ scaling of $\alpha_n$ is in agreement with the trend shown for the doubly clamped beam by the solid line in Fig.~\ref{fig:omegan_over_omega1}(b). This yields the nondimensional Duffing parameter, $ \bar{\alpha}_n \!=\! \frac{\pi^4}{2} n^4$, which is the dashed line in Fig.~\ref{fig:omegan_over_omega1}(b).

\begin{figure}[h!]
\vspace{1cm}
\begin{center}
\includegraphics[width=3.0in]{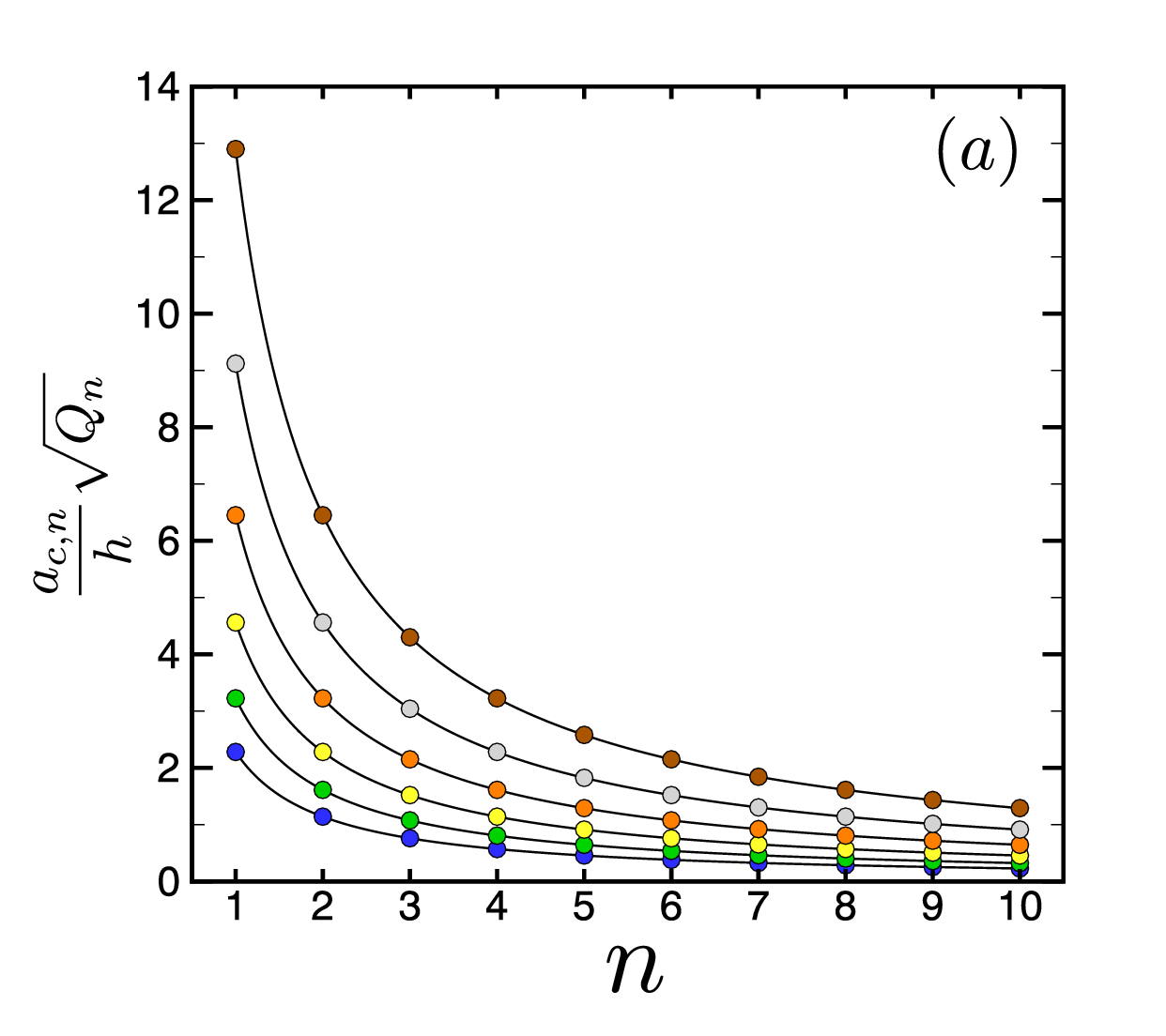}
\includegraphics[width=3.0in]{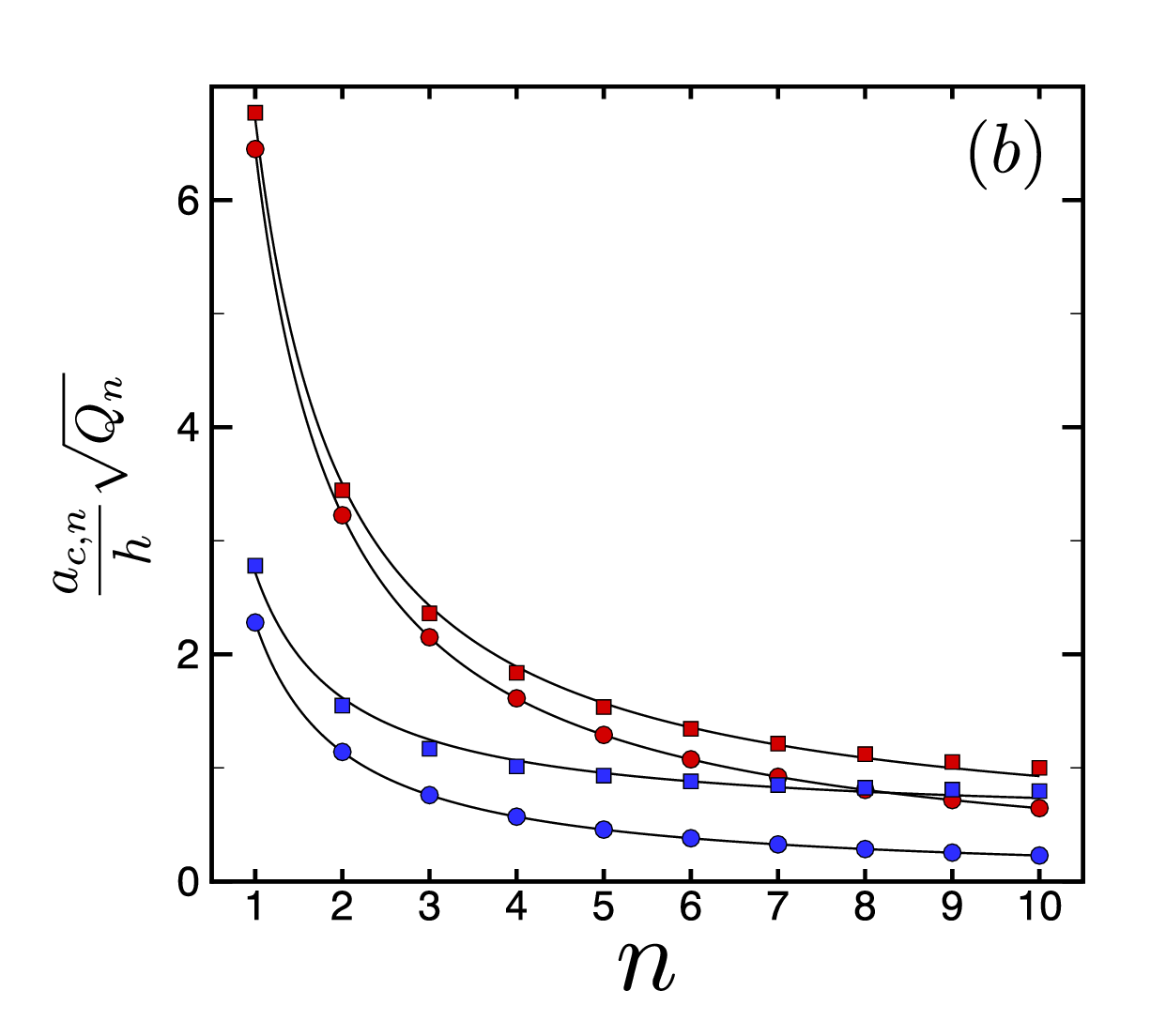}
\end{center}
\caption{(a)~Variation of $a_{c,n} \frac{\sqrt{Q_n}}{h}$ with $n$ for a string as a function of $U$ where (blue) $U\!=\!50$, (green) $U\!=\!100$, (yellow) $U\!=\!200$, (orange) $U\!=\!400$, (gray) $U\!=\! 800$,  (brown) $U\!=\! 1600$. Solid lines are curves through the data using Eq.~(\ref{eq:acnQoverH-string}) which illustrate the  $1/n$ dependence. (b) Comparison of the clamped beam (squares) and string (circles) for $U\!=\!400$ (red) and $U\!=\!50$ (blue).} 
\label{fig:acqoverh-string-clamped}
\end{figure}

Finally, the critical amplitude $a_{c,n}$ is found using $\omega_n$ and $\alpha_n$ to be
\begin{equation}
a_{c,n} = \left( \frac{4}{3^{5/4}} \frac{h}{\pi} \sqrt{\frac{U}{Q_n}} \right) n^{-1}
\label{eq:acn-string}
\end{equation}
which can be expressed as 
\begin{equation}
\frac{a_{c,n}}{h} \sqrt{Q_n} = \left( \frac{4 \sqrt{U}}{3^{5/4}\pi} \right) n^{-1}.
\label{eq:acnQoverH-string}
\end{equation}
Using the notation of Eq.~(\ref{eq:bending-tension}), $\beta\!=\!0$ due to the absence of a bending contribution, and $\tau \!=\! \left( \frac{4 \sqrt{U}}{3^{5/4}\pi} \right)$ where  $\tau/n$ describes the modal dependence of the ratio of intrinsic tension to stretching induced tension.  This clearly illustrates the $n^{-1}$ dependence and also yields a $U^{1/2}$ scaling with the tension parameter. Therefore, for the string model which only includes contributions from tension, the critical amplitude asymptotically approaches zero for large $n$ and for all values of $U$.  The variation of the scaled critical amplitude for the string model is shown in Fig.~\ref{fig:acqoverh-string-clamped}(a) for several values of $U$. The solid lines are given by Eq.~(\ref{eq:acnQoverH-string}).

The $n^{-1}$ trend is evident in the variation of the scaled critical amplitude for a beam with clamped boundaries that is shown in Fig.~\ref{fig:acqoverh-clamped}. However, the critical amplitude for the beam in Fig.~\ref{fig:acqoverh-clamped} levels off to a finite value for large $n$ which is related to the bending contributions that are neglected here for the string.

A direct comparison between the string and clamped-beam results are shown in Fig.~\ref{fig:acqoverh-string-clamped}(b) for a low tension case ($U\!=\!50$, blue) and a high tension case ($U\!=\!400$, red).  The string model is represented as circles and the clamped-beam model is represented as squares. The deviation between the clamped beam and string models becomes more significant as $n$ increases. This is a result of the increasing magnitude of $\phi_n''(x)$ with $n$ which appears in Eq.~(\ref{eq:omegan-bending}). The importance of bending can be illustrated further using a hinged boundary condition for an elastic beam under tension which we discuss in Section~\ref{section:hinged}.

The spring constant for the different modes of a string can be found to be $k_n \!=\! \left( \frac{\pi^2}{2} \frac{F_T}{L} \right) n^2$ where $k_n \sim n^2$. The quadratic variation of $k_n$ with $n$ is shown in Fig.~\ref{fig:kn-over-k1-clamped}(a) by the dashed line. The amplitude of motion for a thermally driven string can be expressed as $a_{th,n} \!=\! \left(\frac{1}{\pi} \sqrt{\frac{2 k_B T L }{F_T}} \right) n^{-1}$ which yields $a_{th,n} \!\sim\! n^{-1}$. The variation of $a_{th,n}$ with $n$ for the string is shown by the dashed line in Fig.~\ref{fig:kn-over-k1-clamped}(b). Inserting the expressions for $a_{c,n}$ and $a_{th,n}$ for the string into Eq.~(\ref{eq:ldr}) and rearranging yields 
\begin{equation}
    \mathcal{L}_n = 0.156 U \sqrt{\frac{E b h^2}{k_B T} } \left(\frac{h}{L} \right)^{3/2} Q_n^{-1/2}.
\label{eq:ldr-string}
\end{equation}
The linear dynamic range of the string has a linear dependence on $U$ where the mode number dependence enters in only through $Q_n$. Therefore, the quantity $\mathcal{L}_n Q_n^{1/2}$ is a mode-independent constant if the geometry and temperature are held fixed.  There is a strong dependence upon the geometry and, in particular, through the inverse aspect ratio $h/L$ where long and thin strings, $h/L \ll 1$, will have a decreased linear dynamic range. 

\subsection{A beam with hinged boundaries}
\label{section:hinged}

A useful description of an elastic beam that is under high tension is to replace the clamped boundaries with hinged (or pinned) boundaries where
$W(0,t) \!=\! W(1,t) \!=\! W''(0,t) \!=\! W''(1,t) \!=\! 0.$
An analysis using the full beam equation with the hinged boundaries includes contributions from bending and tension and  yields insightful closed-form analytical expressions. The hinged description is most useful in the large tension limit which is precisely where the doubly-clamped beam description becomes difficult to use and is difficult to evaluate numerically. 

A hinged boundary can not support a moment and has zero displacement.  The mode shape of a hinged beam is $\phi_n(x) = \sin \left( n \pi x \right)$ which is precisely the same as the string mode shape shown by the blue curves in Fig.~\ref{fig:phi-comp}. However, an important point is that both bending and intrinsic tension contribute to the natural frequency $\omega_n$ of the hinged beam. The bending contribution is found by inserting the mode shape for the hinged beam into Eq.~(\ref{eq:omegan-bending}) to yield $\omega_{n,\beta}^2 \!=\! \left( \frac{E I}{\rho A L^4} \right) n^4 \pi^4$. This yields the scaling $\omega^2_{n,\beta} \!\sim\! n^4$.

The intrinsic tension contribution is found using Eq.~(\ref{eq:omegan-tension}) to yield $\omega_{n,\tau}^2 \!=\! \left( \frac{F_T}{\rho A L^2} \right) n^2 \pi^2$. This is identical to the tension contribution of the string and yields $\omega_{n,\tau}^2 \!\sim\! n^2$. By comparing the $n^4$ scaling for bending with the $n^2$ scaling for intrinsic tension we have the insightful result that the relative importance of bending will increase with increasing mode number. After rearranging, the natural frequency for the hinged beam can be expressed as $\omega_n^2 = \frac{E I}{\rho A L^4} \left( \pi^4 n^4 + 2 \pi^2 U n^2\right)$.

It will be useful to express the ratio of the tension to bending contributions to the frequency as 
$\psi_n \!=\! \omega_{n,\tau}^2/\omega_{n,\beta}^2$ where $\psi_n$ is a nondimensional number relating the significance of intrinsic tension to bending where
\begin{equation}
\psi_n = \frac{U}{\frac{n^2 \pi^2}{2}}.
\label{eq:psin}
\end{equation}
In this light, $\psi_n$ can be recognized as a mode-dependent tension parameter. $\psi_n$ asymptotically approaches zero for $n \! \gg \! 1$ indicating the vanishing role of intrinsic tension for large mode number. $\psi_n$ can be used to define a mode number, $n_c$, where the intrinsic tension contribution is $\psi_n$ times greater than the bending contribution, this can be expressed as $n_c = \frac{1}{\pi} \sqrt{\frac{2 U}{\psi_n}}$.

\begin{figure}[h!]
\vspace{1cm}
\begin{center}
\includegraphics[width=3.5in]{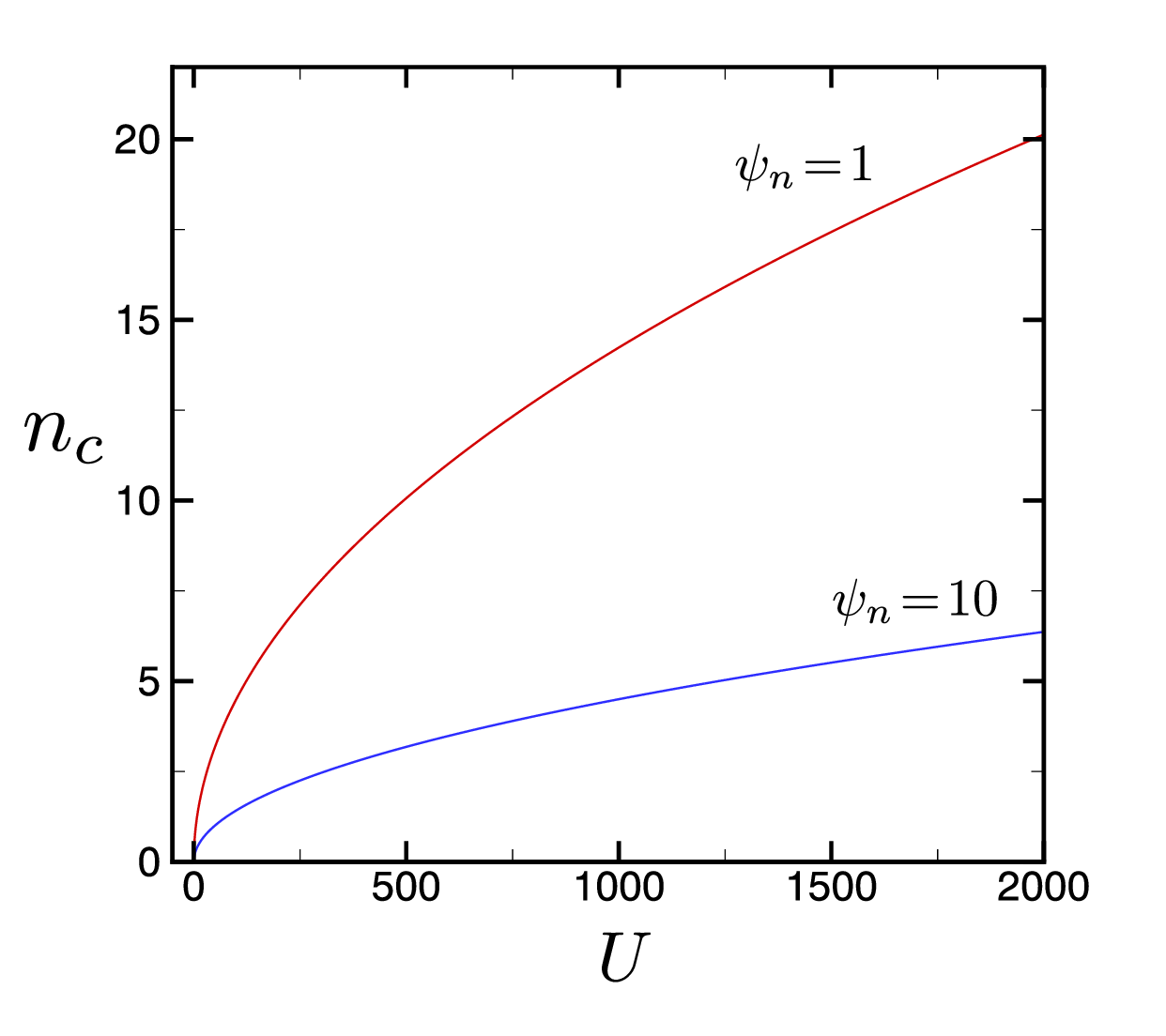}
\end{center}
\caption{Variation of the mode number $n_c$ with $U$ for $\psi_n\!=\!1$ (upper, red) and $\psi_n\!=\!10$ (lower, blue) for an elastic beam with hinged boundaries. For $\psi_n\!=\!1$ the tension and bending contributions are equal and for $\psi_n\!=\!10$ the tension contribution is ten times that of bending.}
\label{fig:nc}
\end{figure}

For a given value of the tension parameter $U$, the tension and bending contributions to the resonant frequency of mode $n$ are equal when $\psi_n \!=\!1$. This is shown in Fig.~\ref{fig:nc} by the upper curve (red). For example, for a low tension case such as $U\!=\!100$ this yields $n_c \!=\! 4.5$, indicating that for modes $n\! \ge \!5$ the bending contribution would be larger than the tension contribution. Alternatively, for a high tension case such as $U\!=\!1000$, this yields $n_c \!=\! 14.24$ where the bending contribution does not become dominant until modes 15 and higher. This quantifies the insightful trend that the role of tension is dominant for more modes as the tension is increased as expected.

A curve for $\psi_n\!=\!10$ is shown in Fig.~\ref{fig:nc} by the lower curve (blue) to indicate the trend for $n_c$ when the tension contribution is ten times that of bending. The variation of $n_c$ with $\psi_n$ provides useful insights into the modal variation of the relative contribution of tension and bending which can be used to guide the theoretical description of the critical amplitude. 

The strength of the Duffing nonlinearity $\alpha_n$ for a hinged beam is identical to the string result. Using this expression along with $\omega_n$ yields the following expression for the scaled critical amplitude for the hinged beam
\begin{equation}
\frac{a_{c,n}}{h} \sqrt{Q_n} = \frac{2 \sqrt{2}}{3^{5/4}} \left(1 + \psi_n \right)^{1/2}.
\label{eq:acn-hinged}
\end{equation}
There are several interesting insights to be drawn from Eq.~(\ref{eq:acn-hinged}). First, in the absence of intrinsic tension, $U\!=\!0$ and therefore $\psi_n\!=\!0$, and we are left only with the contributions due to bending and stretching induced tension. Therefore, when  bending contributions are present and intrinsic tension is absent, this yields a mode-independent constant value for the scaled critical amplitude of $\frac{a_{c,n}}{h} \sqrt{Q_n} \!=\! \frac{2 \sqrt{2}}{3^{5/4}} \!=\! 0.716$.  It is important to highlight that this value is very close to the predicted value of $\frac{a_{c,n}}{h} \sqrt{Q_n} \!=\! 0.739$ using the clamped beam description in the limit of vanishing intrinsic tension discussed in \S\ref{section:clamped}. These two values are not expected to agree exactly due the different mode shapes and boundary conditions that are used in the clamped and hinged models.

For finite values of the tension parameter, $U \!>\! 0$, the contribution due to intrinsic tension asymptotically approaches zero with increasing mode number as indicated by the scaling $\psi_n \!\sim\! n^{-2}$ in Eq.~(\ref{eq:psin}). Therefore, for finite values of $U$, the scaled critical amplitude asymptotically approaches the bending and stretching induced tension value of $\frac{a_{c,n}}{h} \sqrt{Q_n} \!\approx\! 0.716$ with increasing $n$. This is in contrast to the prediction of the string model, which does not include bending, where the scaled critical amplitude vanishes with increasing mode number as indicated by Eq.~(\ref{eq:acnQoverH-string}). Therefore, the hinged model includes the important bending contribution that dominates for large mode number for any beam with finite tension.

\begin{figure}[h!]
\vspace{1cm}
\begin{center}
\includegraphics[width=3.5in]{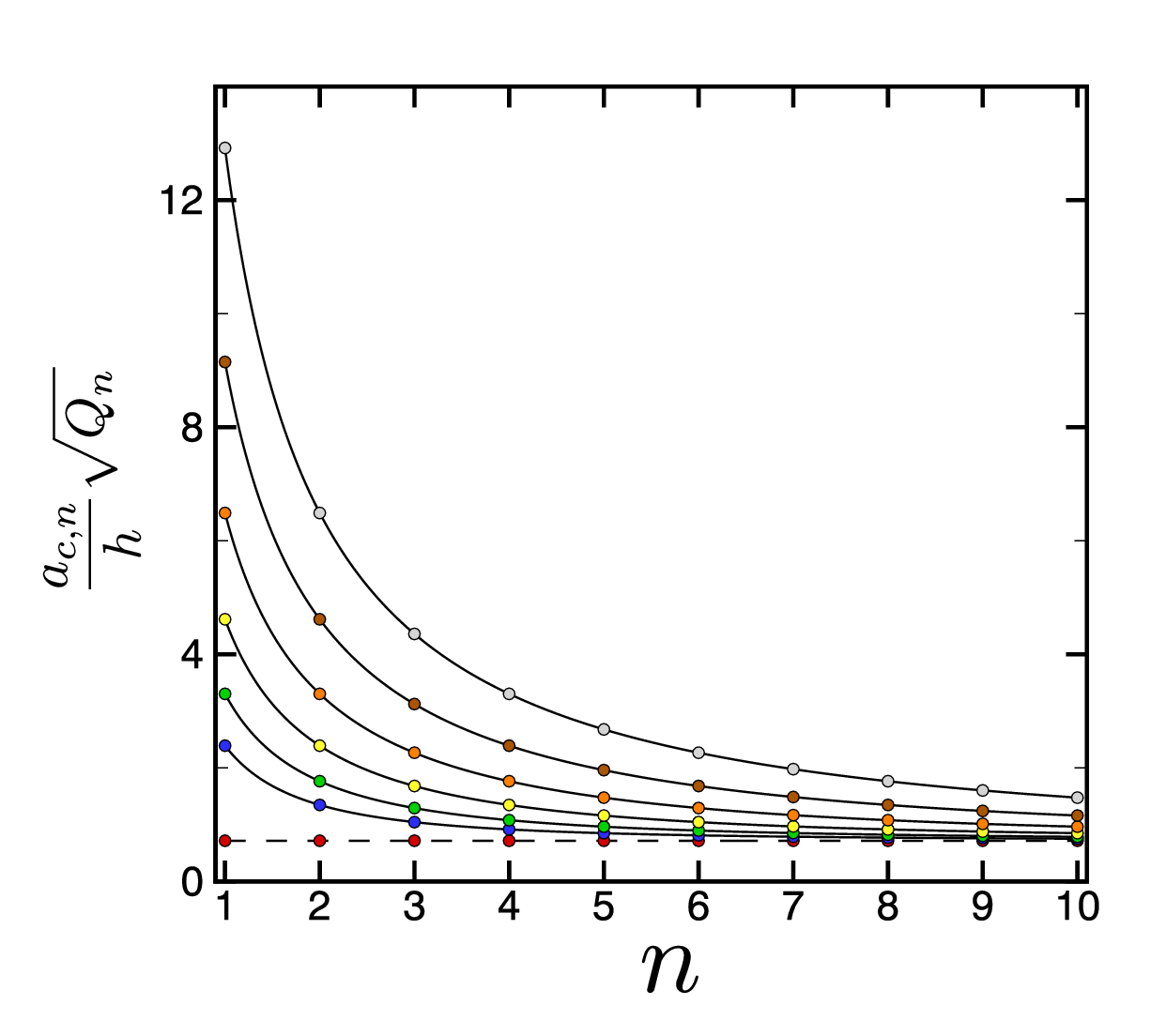}
\end{center}
\caption{Variation of the scaled critical amplitude with mode number for an elastic beam with hinged boundaries given by Eq.~(\ref{eq:acn-hinged}). The dashed line is for the case with only bending, $U=0$, where $\frac{a_{c,n}}{h} \sqrt{Q_n} = \frac{2 \sqrt{2}}{3^{5/4}}\!=\!0.716$. The solid lines are for $U\!=\!50, 100, \!\ldots\!, 1600$ which occur in sequential order from the bottom (blue) to the top (gray). } 
\label{fig:acqoverh-hinged}
\end{figure}

The variation of the scaled critical amplitude with mode number is shown in Fig.~\ref{fig:acqoverh-hinged} for an elastic beam with hinged boundaries. Seven curves are shown for different values of $U$. The horizontal dashed line is the mode-independent value in the absence of intrinsic tension ($U\!=\!0$). The remaining curves, shown with solid lines, are the variation of the scaled critical amplitude for different values of $U$ where $U\!=\!50,100,\!\ldots\!,1600$ which occur in order from bottom (blue) to top (gray).

The trends shown in Fig.~\ref{fig:acqoverh-hinged} yield an insightful description of the roles of bending and tension in the scaled critical amplitude of an elastic beam.  When only bending is present, the result is the constant given by the dashed line which is independent of mode number. The role of  intrinsic tension is to lift the curve from this constant value. The increase in the scaled critical amplitude becomes larger with increasing intrinsic tension and scales as $\frac{\sqrt{U}}{n}$. For example, for the fundamental mode $n\!=\!1$, where this increase is the largest, the increase in the scaled critical amplitude increases as $\sqrt{U}$ which is evident in Fig.~\ref{fig:acqoverh-hinged}. However, the intrinsic tension contribution asymptotically vanishes with a $n^{-1}$ scaling for a fixed value of $U$. This is indicated by all curves in Fig.~\ref{fig:acqoverh-hinged}, for finite $U$, asymptotically decaying to a constant value with increasing $n$.

\begin{figure}[h!]
\vspace{1cm}
\begin{center}
\includegraphics[width=3.2in]{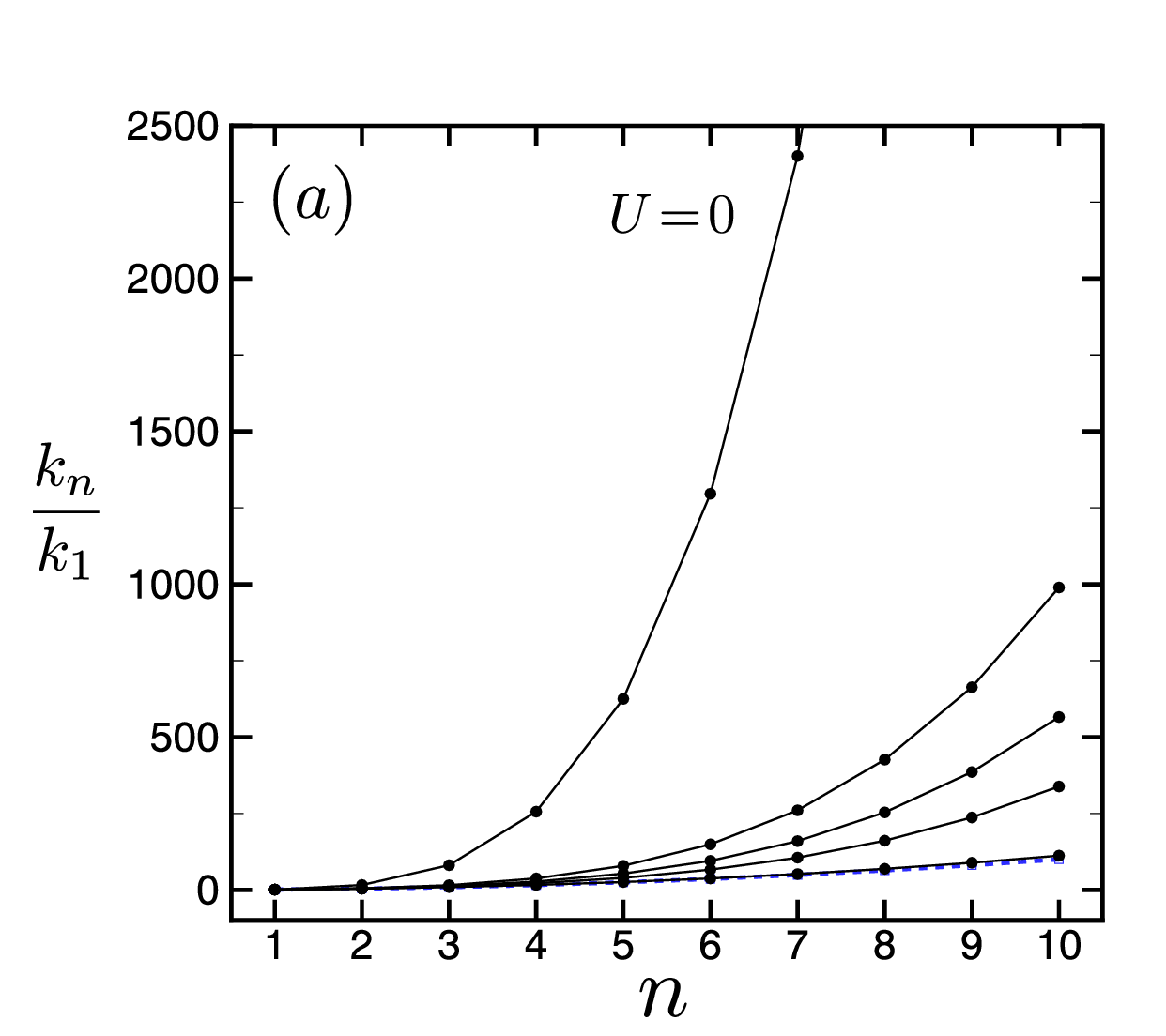}
\includegraphics[width=3.2in]{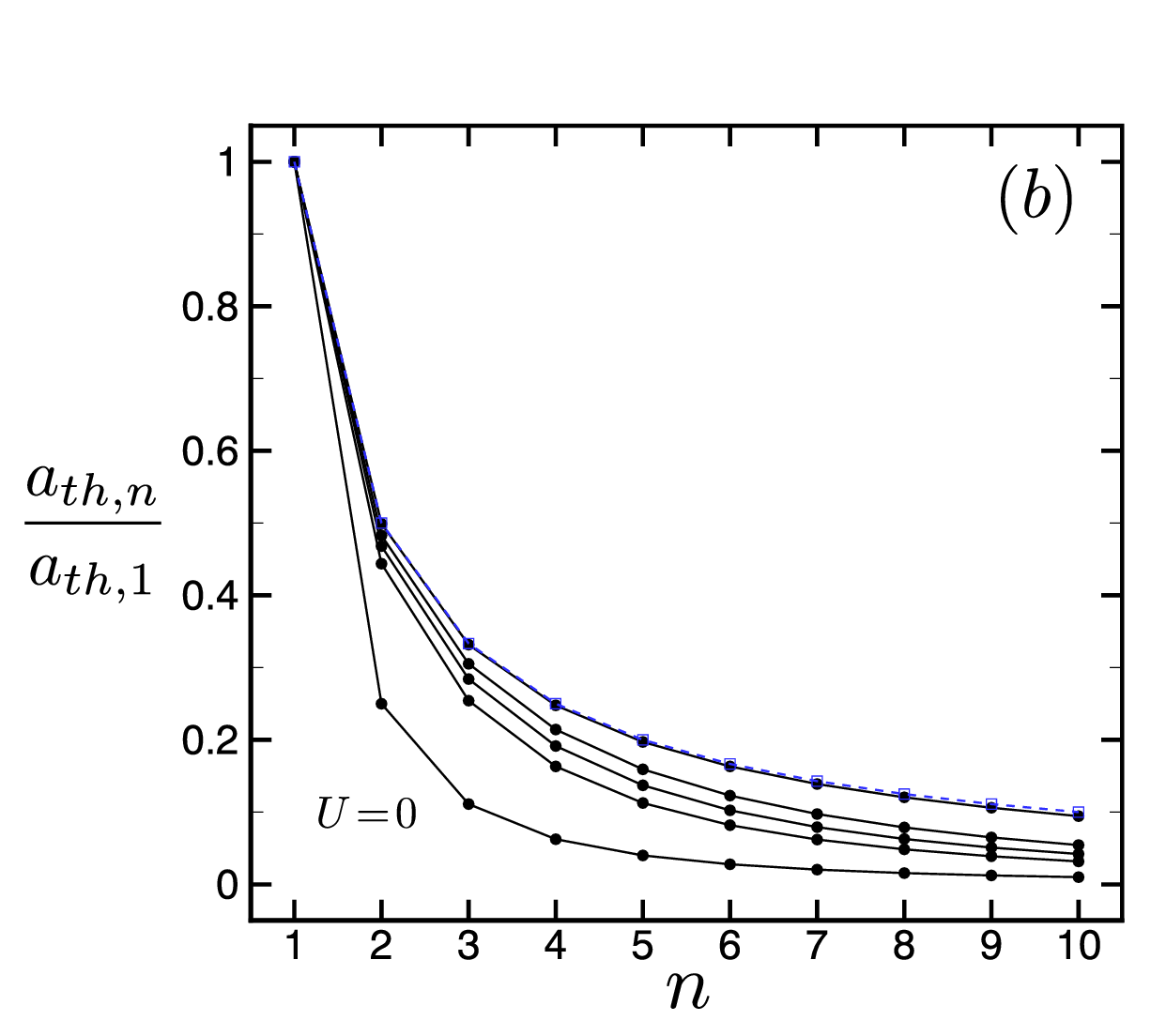}
\end{center}
\caption{Variation of the spring constant~($a$) and thermal amplitude of oscillation~($b$) for a beam with hinged boundaries. ($a$) Solid lines show variation of $k_n$. Results for $U\!=\!0$ are labeled, the remaining solid curves are for $U \!=\! 50, 100, 200$ as the dashed line (blue), representing the results for a string, are approached.} 
\label{fig:kn-athn-hinged}
\end{figure}

The effective mass, $m_n$, of the hinged beam is identical to that of the spring with $m_n\!=\!m/2$ for all $n$. The modal spring constant $k_n$ of the hinged beam can be found using the modal mass and the natural frequency to yield $k_n \!=\! \frac{E I}{2 L^3} \left( \pi^4 n^4 + 2 \pi^2 n^2 U\right)$.
For $U \!\ll\! 1$, the spring constant is dominated by bending and scales as $k_n \sim n^4$. This is illustrated in Fig.~\ref{fig:kn-athn-hinged}(a) by the $U\!=\!0$ curve. However, when intrinsic tension dominates, $U \!\gg\! 1$, this yields $k_n \!\sim\! n^2$. This is illustrated in Fig.~\ref{fig:kn-over-k1-clamped} by the $U\!=\!4000$ curve. This high tension curve overlaps with the string result shown by the dashed line (blue) in Fig.~\ref{fig:kn-athn-hinged}(a).

The amplitude of oscillations due to thermal motion can be determined using $k_n$ and the equipartition of energy. The amplitude of thermal motion for the hinged beam can be expressed as 
\begin{equation}
    \frac{a_{th,n}}{h} = \left( \frac{2 k_B T}{E \, b \, h^2 } \right)^{1/2} \left( \frac{L}{h} \right)^{3/2} \left( n^4 \pi^4 + 2 n^2 \pi^2  U \right)^{-1/2}.  
    \label{eq:athn-hinged}
\end{equation}
The amplitude of thermal motion asymptotically approaches zero as $n$ becomes large. 
The variation of the thermal amplitude for the hinged beam is shown in Fig.~\ref{fig:kn-athn-hinged}(b) for several values of the tension parameter $U$. From Eq.~(\ref{eq:athn-hinged}) it is clear that the thermal amplitude of motion increases for long and thin beams with increasing aspect ratio $L/h$. In the limit of small intrinsic tension, $U \!\ll\! 1$, this yields that the thermally driven amplitude decreases with increasing mode number as $a_{th,n} \!\sim\! n^{-2}$. This is shown in Fig.~\ref{fig:kn-athn-hinged}(b) by the curve labeled $U=0$. For very large values of intrinsic tension, $U \gg 1$, the thermal amplitude scales as $a_{th,n} \!\sim\! n^{-1}$. The $n^{-1}$ scaling of $a_{th,n}$ is shown by the curve for $U\!=\!4000$ in Fig.~\ref{fig:kn-athn-hinged}(b). The high intrinsic tension prediction overlaps with the string result shown by the blue dashed curve.

An expression for the linear dynamic range of a hinged beam under tension can be determined using Eqs.~(\ref{eq:acn-hinged})-(\ref{eq:athn-hinged}) in Eq.~(\ref{eq:ldr}). After some rearranging, this yields
\begin{equation}
    \mathcal{L}_n = \frac{0.754}{3^{5/4}} \left( \frac{E b h^2}{k_B T} \right)^{1/2} \left( \frac{h}{L} \right)^{3/2} \left( \frac{n^2 \pi^2 + 2 U}{\sqrt{Q_n}} \right).
\end{equation}
We will explore this result in detail for two specific beams in Section~\ref{section:comparison}.

\subsection{Insights from the Clamped Beam, String, and Hinged Beam Models}
\label{section:insights}

Our overarching goal is to theoretically describe the mode-dependent linear dynamic range of micro and nanoscale oscillators composed of elastic beams that are under tension. We have used three different theoretical approaches to describe this system: a doubly-clamped elastic beam, a string, and a doubly-hinged elastic beam. Each approach has advantages and disadvantages.

From a physical point of view, the most appropriate model to use is an elastic beam with clamped boundaries as described in \S\ref{section:clamped}.  This model includes bending and tension and is valid for all values of the tension parameter $U \!\ge\! 0$. However, its mathematical description is quite complicated and does not yield insightful closed-form analytical expressions. This is most evident for very high tension cases, $U \gtrsim 10^3$, where the expressions require care in their numerical evaluation due to hyperbolic functions with large arguments and the sharp features in the mode shapes near the boundaries as indicated in Fig.~\ref{fig:phi-comp}. As we showed in \S\ref{section:clamped}, the expressions can be evaluated numerically to yield important insights.

A very useful and insightful simplification is given by the string model described in \S\ref{section:string}. In this case, bending is neglected entirely and the role of tension is included. This model immediately yields useful closed-form expressions. The string model is most useful for the description of beams under significant tension when only the dynamics of the modes for small to moderate $n$ are of interest. The limitation in the range of applicability of the string model can be traced to its neglect of the bending contributions which eventually becomes important at higher mode numbers as shown in Fig.~\ref{fig:nc}.

The hinged-beam model developed in \S\ref{section:hinged} includes bending and tension while also yielding insightful closed-form analytical expressions. The clarity of the hinged model, and its analytical description, are its most advantageous qualities. However, its range of applicability is limited to situations with a significant amount intrinsic tension where replacing the clamped boundary with a hinged boundary is appropriate.

In the following, we consider the clamped elastic beam description as the most accurate result and compare with the string and hinged models to assess their usefulness for describing the mode-dependent linear dynamic range of micro and nanoscale beams. We quantify these ideas further by directly comparing the clamped beam, string, and hinged beam models with experimental measurement in \S\ref{section:comparison}.

\begin{figure}[h!]
\vspace{1cm}
\begin{center}
\includegraphics[width=3.0in]{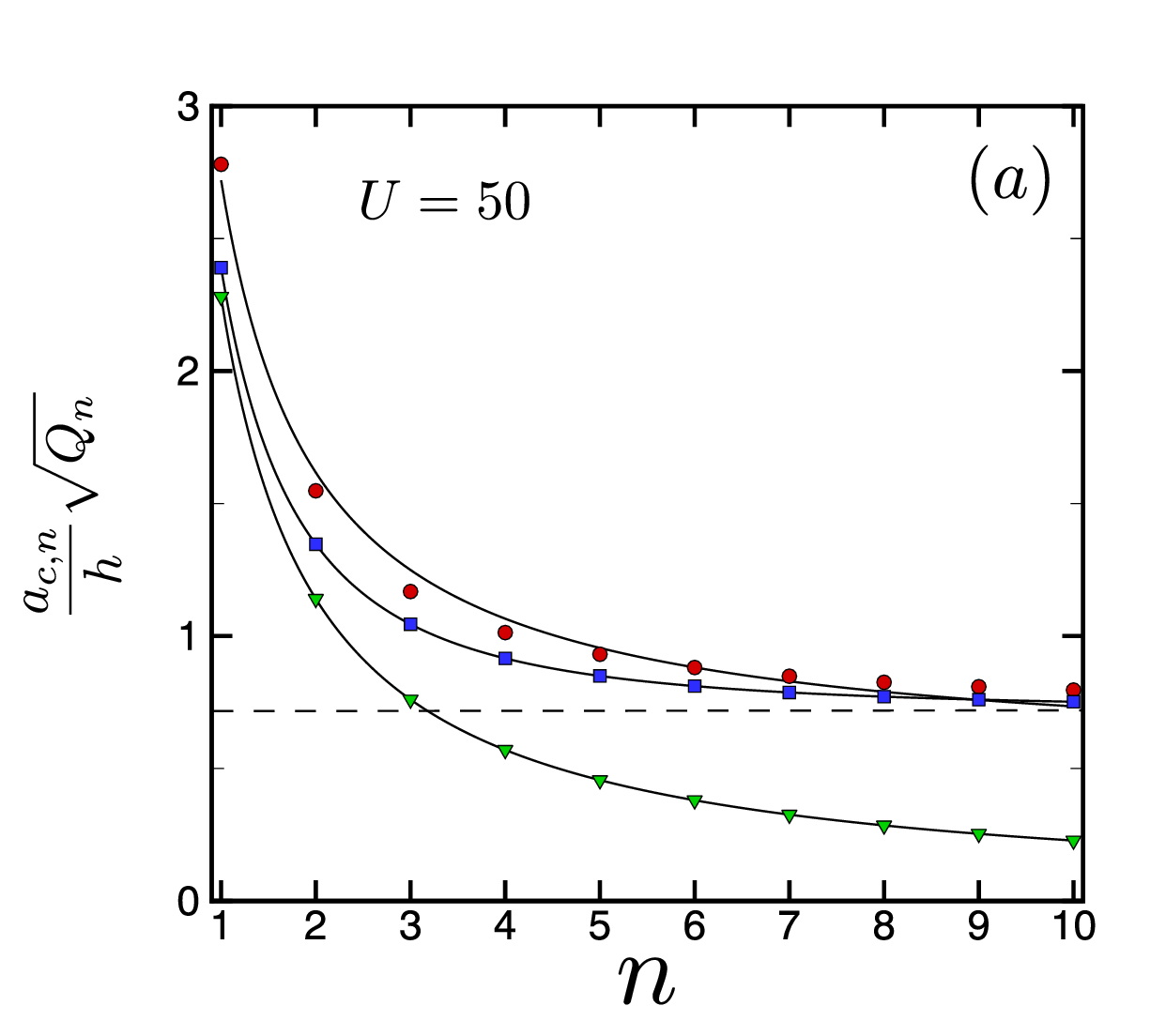}
\includegraphics[width=3.0in]{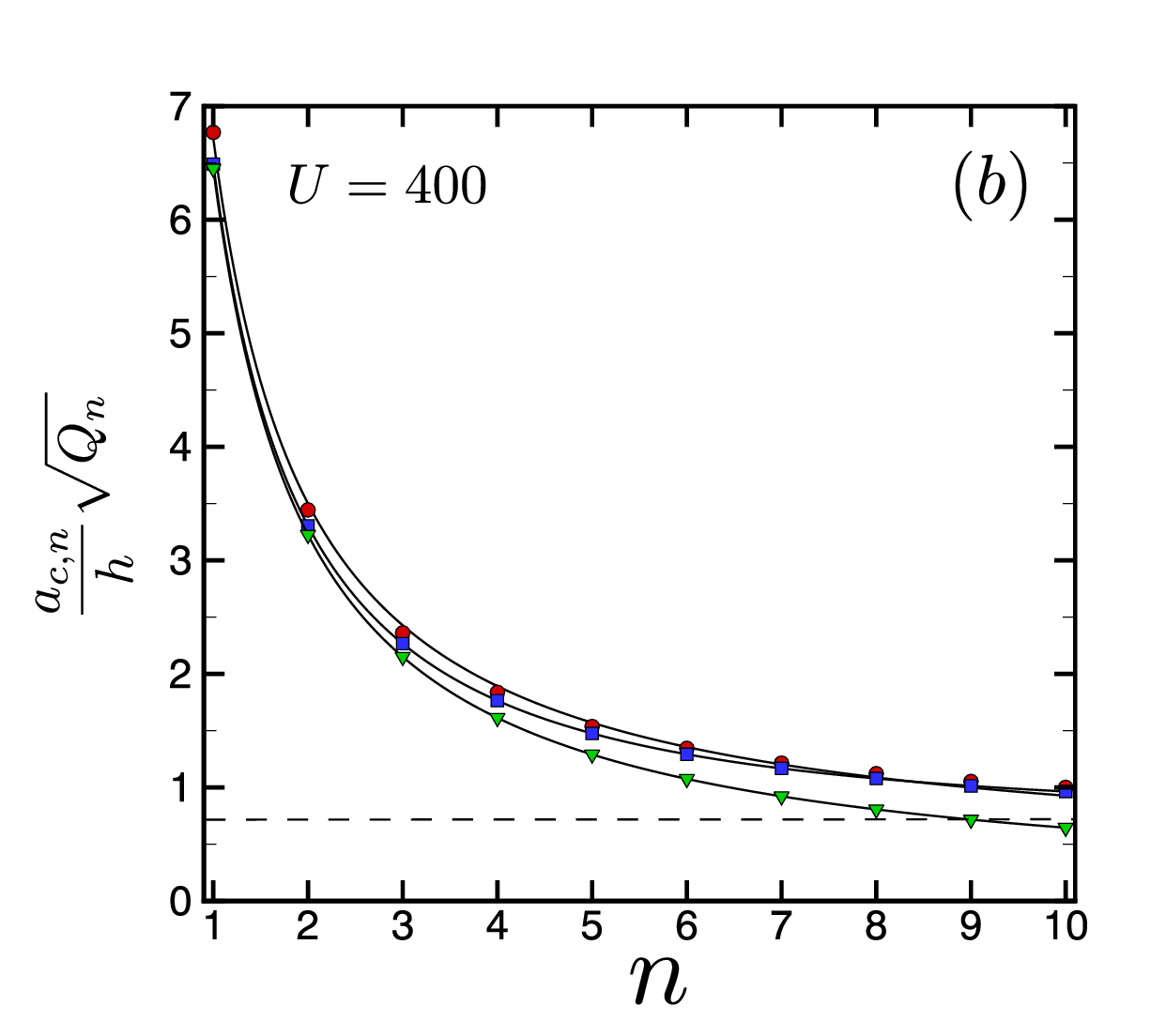}
\end{center}
\caption{Comparison of the scaled critical amplitude for a clamped beam (red, circles), hinged beam (blue, squares), and a string (green, triangles). The dashed line is the large mode number limit of the hinged model, $\frac{a_{c,n}}{h} \sqrt{Q_n} \!=\! \frac{2 \sqrt{2}}{3^{5/4}}$, where only bending contributes. (a)~$U\!=\!50$. (b)~$U\!=\!400$.} 
\label{fig:acqoverh-hinged-comp}
\end{figure}

Figure~\ref{fig:acqoverh-hinged-comp} shows a comparison of the variation of the scaled critical amplitude with mode number for the clamped-beam (circles, red), string (triangles, green), and hinged-beam (squares, blue) models for a low tension case~($a$) and a higher tension case~($b$). In both panels, the horizontal dashed line indicates the bending-only result prediction from the hinge model where Eq.~(\ref{eq:acn-hinged}) reduces to $\frac{a_{c,n}}{h} \sqrt{Q_n} \!\approx\! 0.716$ in the absence of tension where $U\!=\!0$. 

The low tension case shown in Fig.~\ref{fig:acqoverh-hinged-comp}(a) is for $U\!=\!50$. The upper curve (circles, red) are for the clamped beam where the solid line is a curve fit through the data of the form given by Eq.~(\ref{eq:bending-tension}). The results for the clamped boundaries are well described by the $n^{-1}$ trend which approaches the bending only asymptotic value indicated by the dashed line. The prediction of the string model, given by Eq.~(\ref{eq:acnQoverH-string}) and shown by the green triangles, asymptotically approaches a vanishing value for large mode number. By the third mode, there is significant error in the string model. This clearly indicates that the string model does not describe the low tension case well.

The hinged beam model is shown in Fig.~\ref{fig:acqoverh-hinged-comp}(a) by the blue squares where the solid line is given by Eq.~(\ref{eq:acn-hinged}). The error in the hinged model, when compared to the clamped model, for the first several modes is smaller than that of the string model yet the deviations are significant. However, for larger mode number the hinged model approaches the result of the clamped model and the error is small for $n \gtrsim 7$. Overall, even for this low tension case the hinged model provides very useful and insightful results for all $n$ with the error reducing with larger values of $n$.

In Fig.~\ref{fig:acqoverh-hinged-comp}(b), we show the same comparison for the high tension case where $U\!=\!400$. In this case, the string model is much more accurate and remains useful for more modes than in the low tension case. However, by $n\approx10$, the error in the string approximation is becoming significant and will become larger with increasing $n$. The hinged beam description, on the other hand, provides an excellent description for all of the modes shown.

\begin{figure}[h!]
\vspace{1cm}
\begin{center}
\includegraphics[width=3.0in]{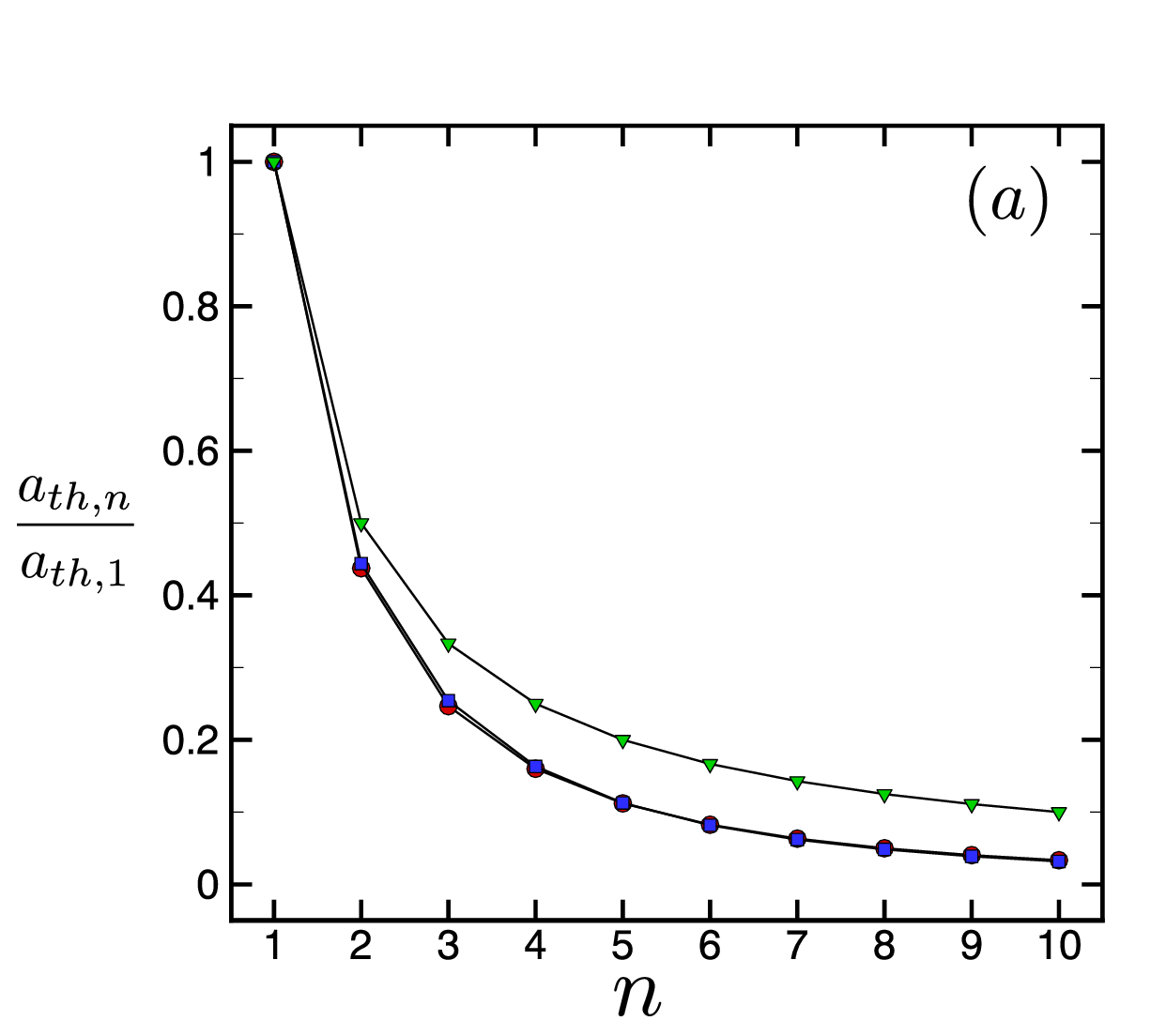}
\includegraphics[width=3.0in]{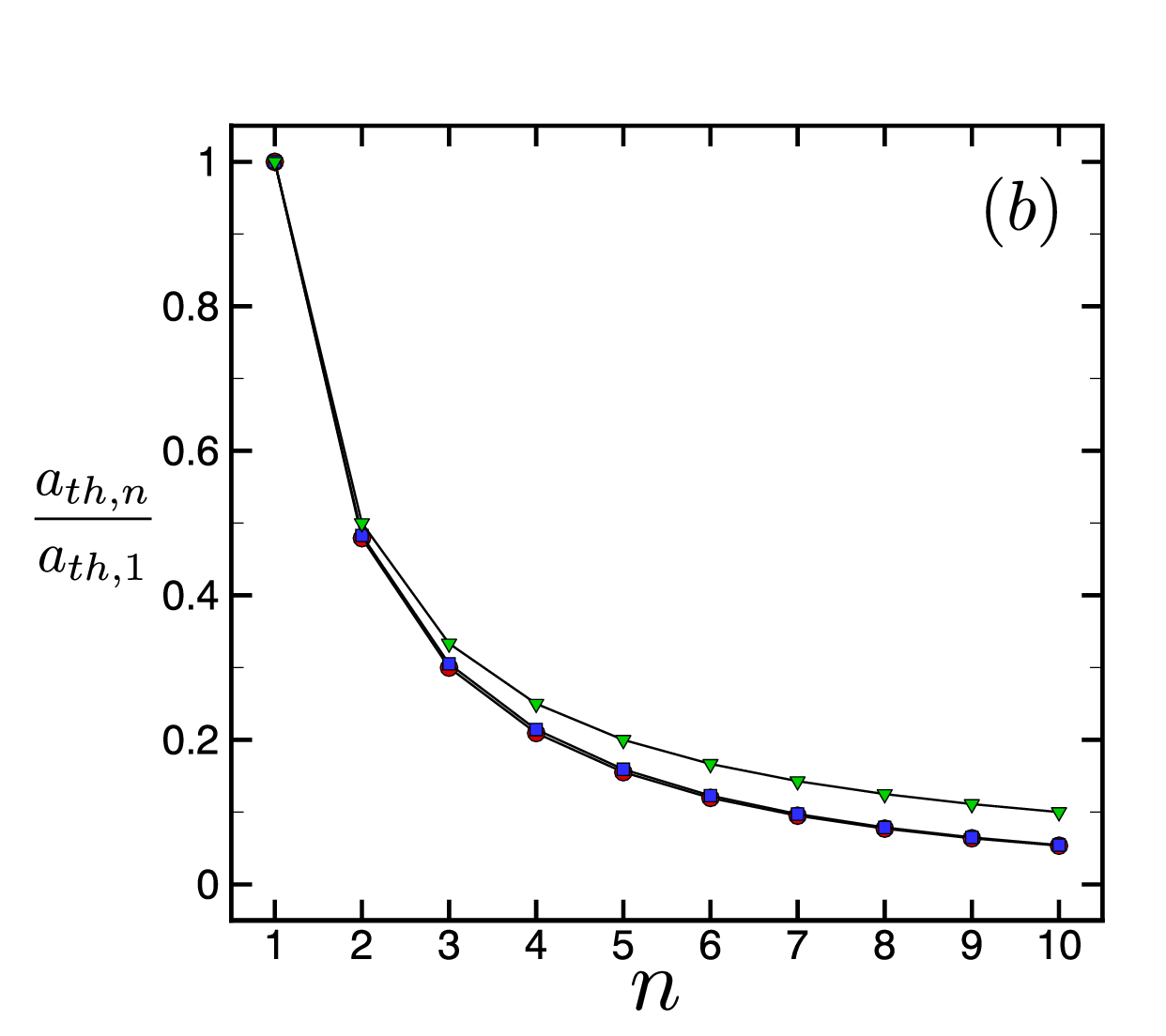}
\end{center}
\caption{A comparison of the scaled thermal amplitude for a clamped beam (red, circles), hinged beam (blue, squares), and a string (green, triangles). (a)~$U\!=\!50$, (b)~$U\!=\!400$.} 
\label{fig:athn-comp}
\end{figure}

A comparison of the amplitude of thermal motion for the different descriptions are shown in Fig.~\ref{fig:athn-comp} for a low tension case $U\!=\!50$~(a) and a high tension case $U\!=\!200$~(b). The hinged approach (squares, blue) is in excellent agreement with the clamped boundary result (circles, red). As the tension increases, the hinged and clamped results continue to approach the string result (triangles, green).

\section{Comparison with Experiment}
\label{section:comparison}

We now compare the theoretical predictions with experimental measurements for two silicon nitride nanobeams which have previously been described using only the string model~\cite{ma:2024}. The properties of the beams are given in Table~\ref{table:beam}. Both beams have the same cross-sectional area where beam 2 is longer than beam 1. Both beams have identical values of the Young's modulus $E$ and density $\rho$. The amount of intrinsic tension, $F_T$, is a result of the fabrication process.  The longer beam, beam 2, is under considerably more tension than the shorter beam, beam 1. This is evident in the values of the tension parameter where $U_2 \!\gg\! U_1$. The values lister for $E$, $\rho$, and $F_T$ have all been determined experimentally~\cite{ma:2024}.
\begin{table}[h!]
    \centering
    \begin{tabular}{c@{\hskip 1cm} c@{\hskip 1cm} c@{\hskip 1cm} c@{\hskip 1cm} c@{\hskip 1cm} c@{\hskip 1cm}}
  \text{Beam} &       $L$        & $b$    & $h$    & $F_T$  & $U$ \\ 
  &       ($\mu$m) & (nm) & (nm) & ($\mu$N) & - \\ \hline
   1 &      30       & 900  & 100  & 8.9    & 213.6 \\ 
   2 &      50       & 900  & 100  & 63.5   & 4233.3  
    \end{tabular}
    \caption{Properties of the two beams used in experiment with length $L$, width $b$, thickness $h$, intrinsic tension force $F_T$, and tension parameter $U$. The beams are made of silicon nitride with a Young's modulus of $E\!=\!250$ GPa and density of $\rho\!=\!3000$ kg/m$^3$ and are at room temperature $T=300$K. }
    \label{table:beam}
\end{table}

\begin{figure}[h!]
\vspace{1cm}
\begin{center}
\includegraphics[width=3.0in]{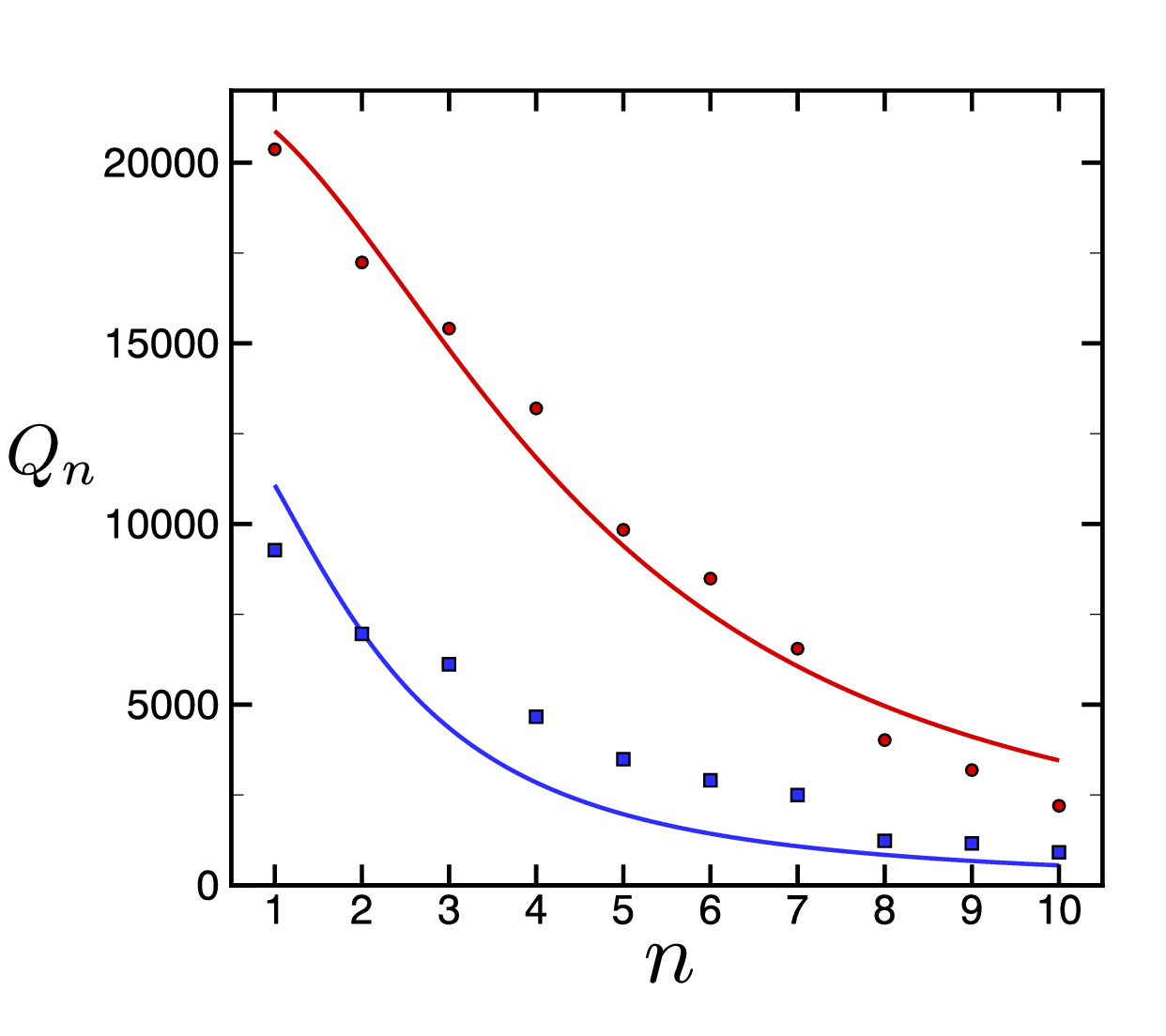}
\end{center}
\caption{Experimental measurements of the variation of the quality factor $Q_n$ with mode number $n$ for the two different beams of Table~\ref{table:beam} in vacuum. (blue, squares) The lower tension beam, $L\!=\!30 ~\mu\text{m}$, $U\!=\!213.6$, $Q_i\!=\!1328.7$. (red, circles) The high tension beam, $L\!=\!50 ~\mu\text{m}$, $U\!=\!4233.3$, $Q_i\!=\!477.5$. The solid lines are curve fits through the data using Eq.~(\ref{eq:qn-dissipation-dilution}) with $Q_i$ as a free parameter.} 
\label{fig:qn}
\end{figure}

The high quality factors of silicon nitride nanobeams, in vacuum and under high intrinsic tension, is a topic of intense interest~\cite{engelsen:2024}.  Measurements of $Q_n$ for the two beams are shown in Fig.~\ref{fig:qn} where the squares (blue) are for beam 1 and the circles (red) are for beam 2. The quality factors of the fundamental mode for both beams is on the order of $10^4$ which then rapidly decays with increasing mode number. Overall, the quality factors of the longer and higher tension beam, beam 1, are much larger than those of beam 2. 

It is now understood that the quality factors of small beams under high tension increase dramatically due to dissipation dilution~\cite{engelsen:2024}. In essence, this is because the elastic energy of a mode of oscillation includes both internal contributions as well as an external contribution due to the intrinsic tension whereas the dissipation is mostly due to internal losses~\cite{schmid:2011,bachtold:2022}. The variation of $Q_n$ can be expressed as~\cite{engelsen:2024}
\begin{equation}
Q_n \!=\! Q_i \left( \sqrt{\frac{2}{U}} + \frac{n^2 \pi^2}{2 U} \right)^{-1}
\label{eq:qn-dissipation-dilution}
\end{equation}
where $Q_i$ is the intrinsic value of the quality factor of the fundamental mode in the absence of imposed tension. The solid curves in Fig.~\ref{fig:qn} are fits through the data using this expression with $Q_i$ as a free parameter.

\begin{figure}[h!]
\vspace{1cm}
\begin{center}
\includegraphics[width=3.0in]{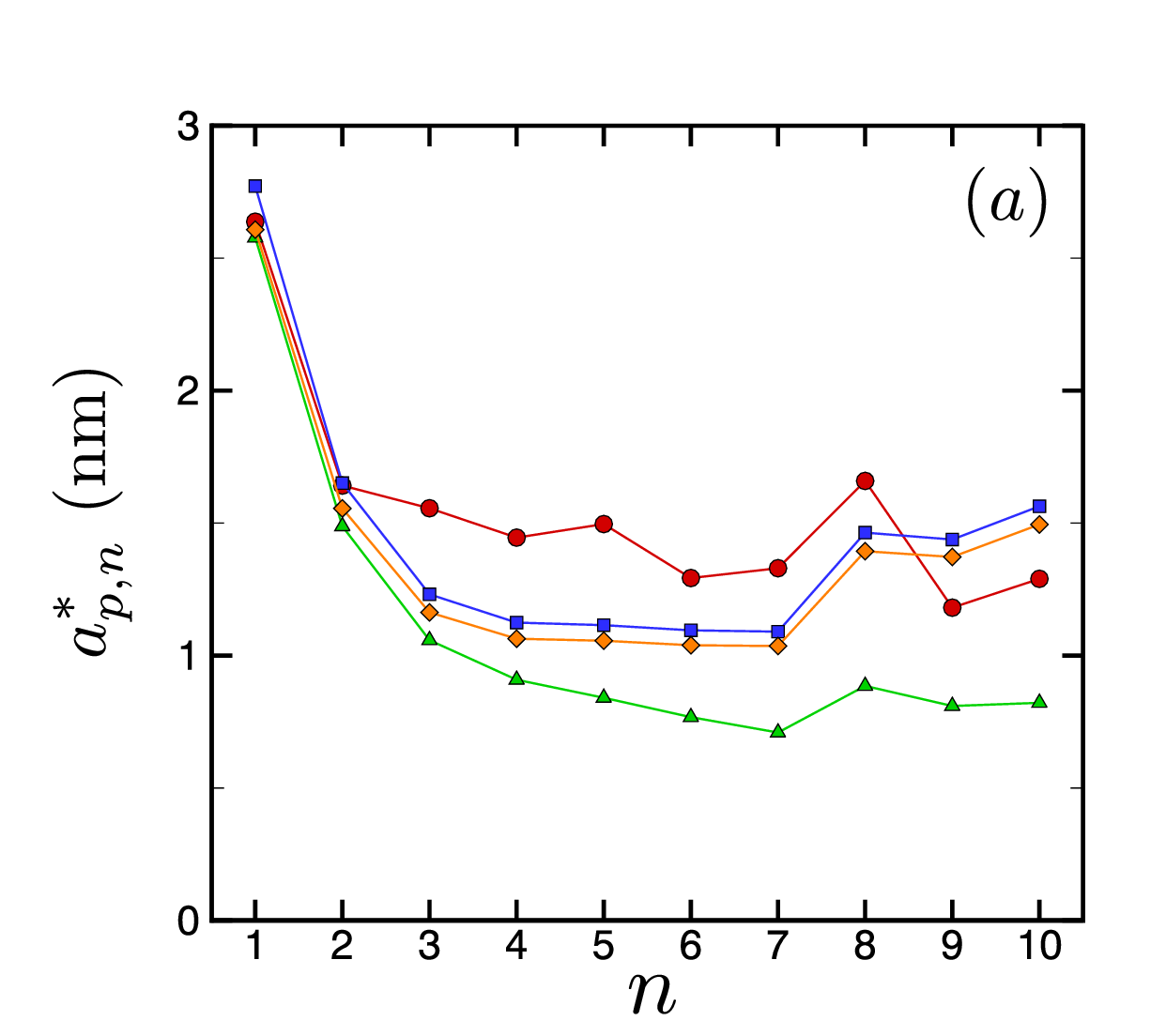}
\includegraphics[width=3.0in]{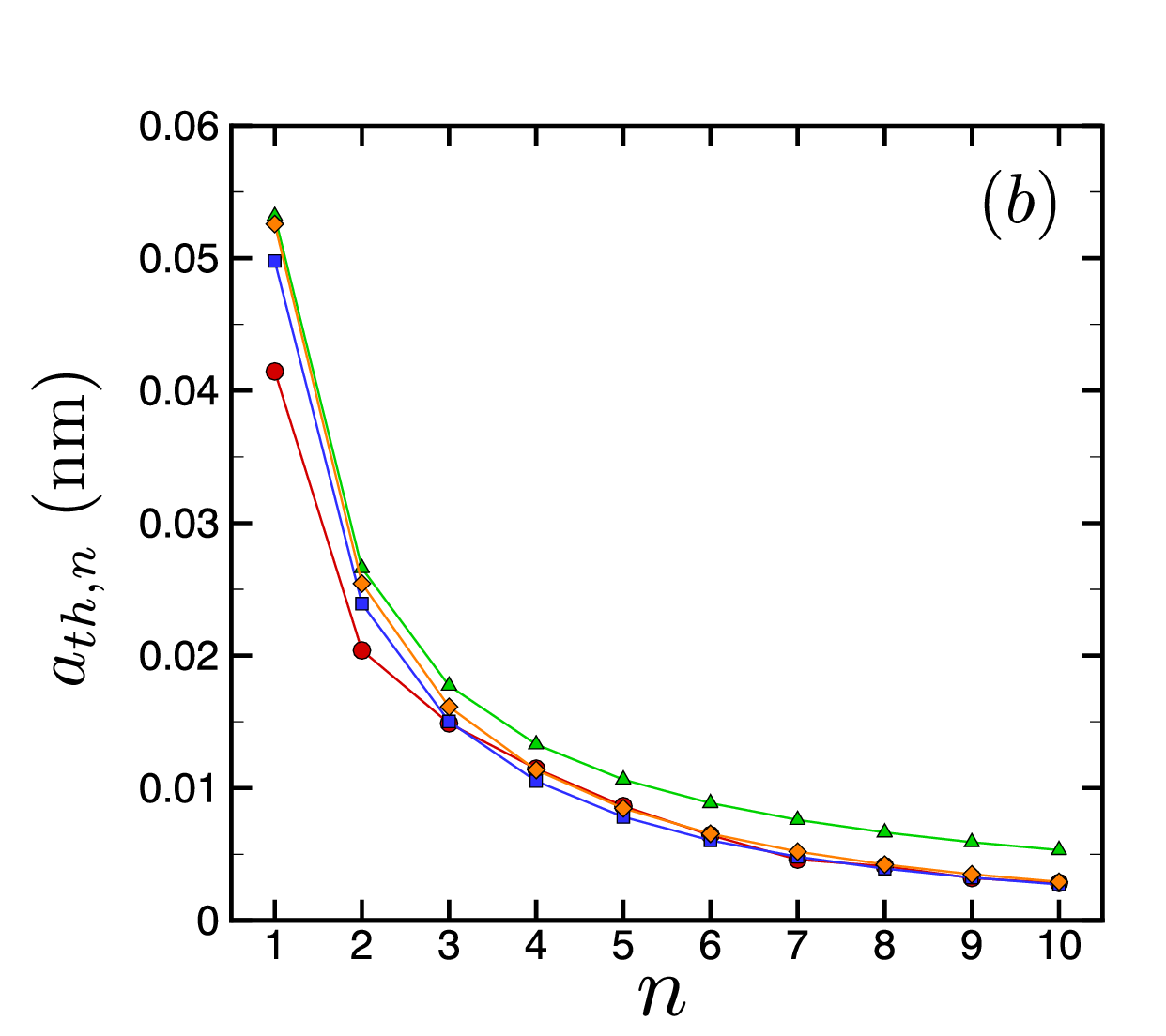}
\includegraphics[width=3.0in]{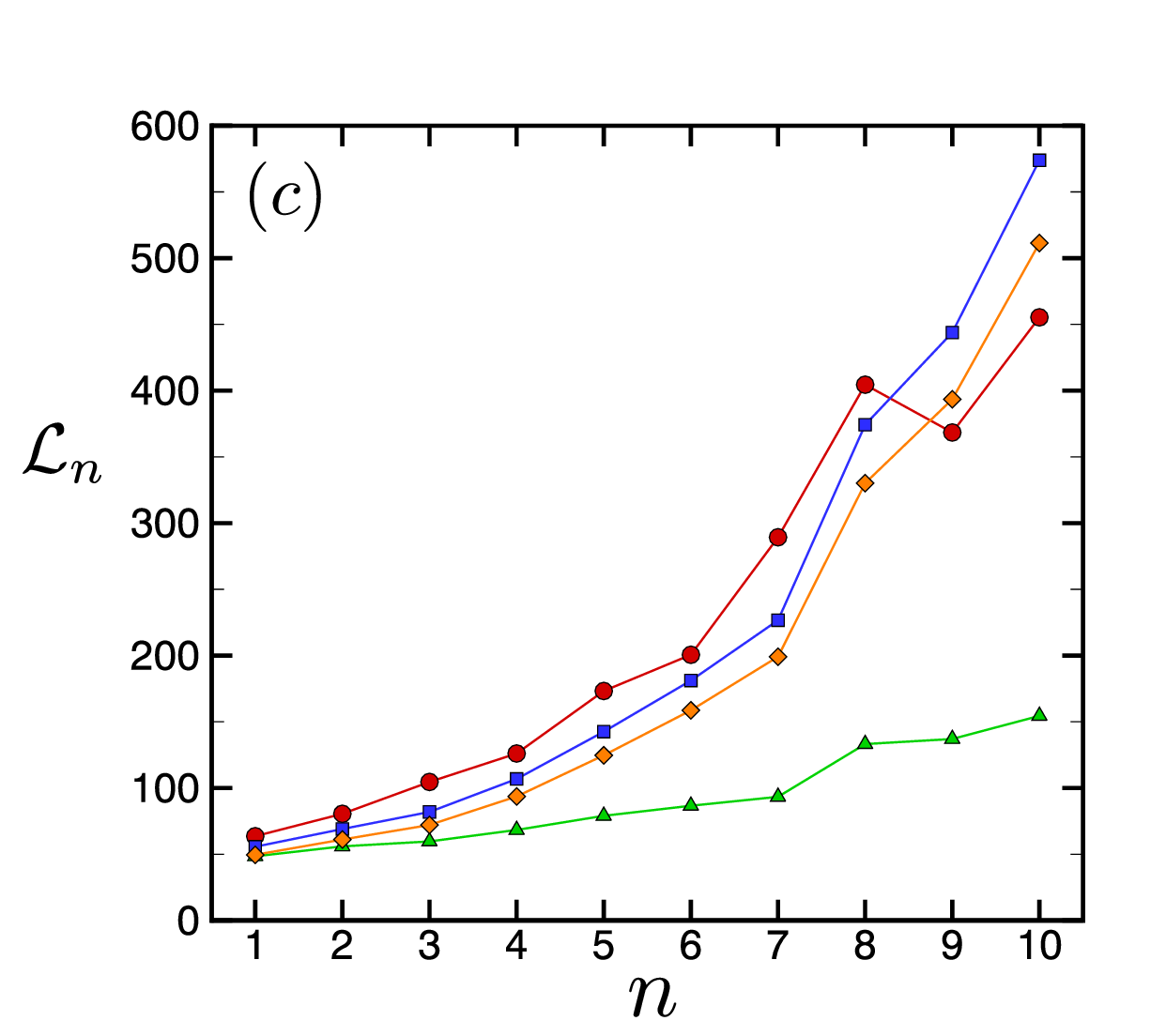}
\end{center}
\caption{Comparison between theory and experiment for the lower tension beam in Table~\ref{table:beam} with $U\!=\!214$. The variation of the ($a$) critical amplitude $a^*_{p,n}$, ($b$) the thermal amplitude $a_{th,n}$, and the ($c$) linear dynamic range $\mathcal{L}_n$. Circles (red) are measurements from experiment. Theoretical predictions are shown for a clamped beam (squares, blue), a string (triangles, green), and a hinged beam (diamonds, orange).} 
\label{fig:beam-U214}
\end{figure}

The comparison between theory and experiment for beam 1 is shown in Fig.~\ref{fig:beam-U214}. In all panels, red circles are the experimental measurements. Theoretical predictions are shown by the blue squares for a clamped beam, the green triangles for a string, and the orange diamonds for a hinged beam. In presenting numerical values for the theoretical predictions we have used the quality factors $Q_n$ from experiment shown in Fig.~\ref{fig:qn}. The non-smooth nature of the theoretical predictions is due to the use of the experimentally measured quality factor as an input for these calculations.

Figure~\ref{fig:beam-U214}(a) shows the variation of the critical RMS amplitude of driven motion, $a^*_{p,n}$, where the nonlinear contribution to the dynamics from stretching induced tension becomes important. The critical value, measured in experiment, for the fundamental mode is $a^*_{p,1} \!=\! 2.63$ nm which yields an amplitude of motion that is less than 3\% of the beam's thickness for reference. The theoretical prediction, using the clamped beam description is shown by the blue squares. The clamped beam prediction is computed using Eq.~(\ref{eq:acn2}) in the expression for $a_{p,n}^*$ with the mode shapes given by Eq.~(\ref{eq:phi-clamped-clamped}) where $Q_n$ are from experiment. 

The green triangles of Fig.~\ref{fig:beam-U214}(a) are the predictions using the string description  where Eq.~(\ref{eq:acn-string}) is used to determine $a_{p,n}^*$. As expected, the string prediction is good for the first several modes, $n \!\lesssim\! 2$, the errors then become significant for larger modes due to the neglect of the bending contribution. The orange diamonds represent the theoretical prediction using the hinged beam description. In this case, $a^*_{p,n}$ is determined using Eq.~(\ref{eq:acn-hinged}). The hinged beam description remains quite close to the clamped beam results for all ten modes that are shown.

The RMS amplitude of thermally driven motion $a_{th,n}$ is shown in Fig.~\ref{fig:beam-U214}(b). The amplitude of thermal motion decreases rapidly with increasing mode number $n$ which can be traced back to the stiffening of the higher modes. The amplitude of thermal motion is smaller than the critical amplitude, $a_{th,n} \!<\! a^*_{p,n}$, as expected, and it is this difference that leads to the linear dynamic range. The clamped beam and hinged beam descriptions are very close, and their overall agreement with experiment is excellent for $n \!\gtrsim\! 3$. The string description exhibits deviations due to its neglect of the bending contribution to the modal spring constants.

The linear dynamic range, $\mathcal{L}_n$, is shown in Fig.~\ref{fig:beam-U214}(c). $\mathcal{L}_n$ increases with increasing mode number. The agreement between the clamped beam description and the experimental measurements is very good. As expected, the string is a good description for several modes and then exhibits increasing deviations from experiment for larger mode numbers. The hinged beam is in very good agreement with the clamped beam and the experiments. The ability of the hinge description to provide an accurate closed-form analytical description is a significant advantage. 

\begin{figure}[h!]
\vspace{1cm}
\begin{center}
\includegraphics[width=3.0in]{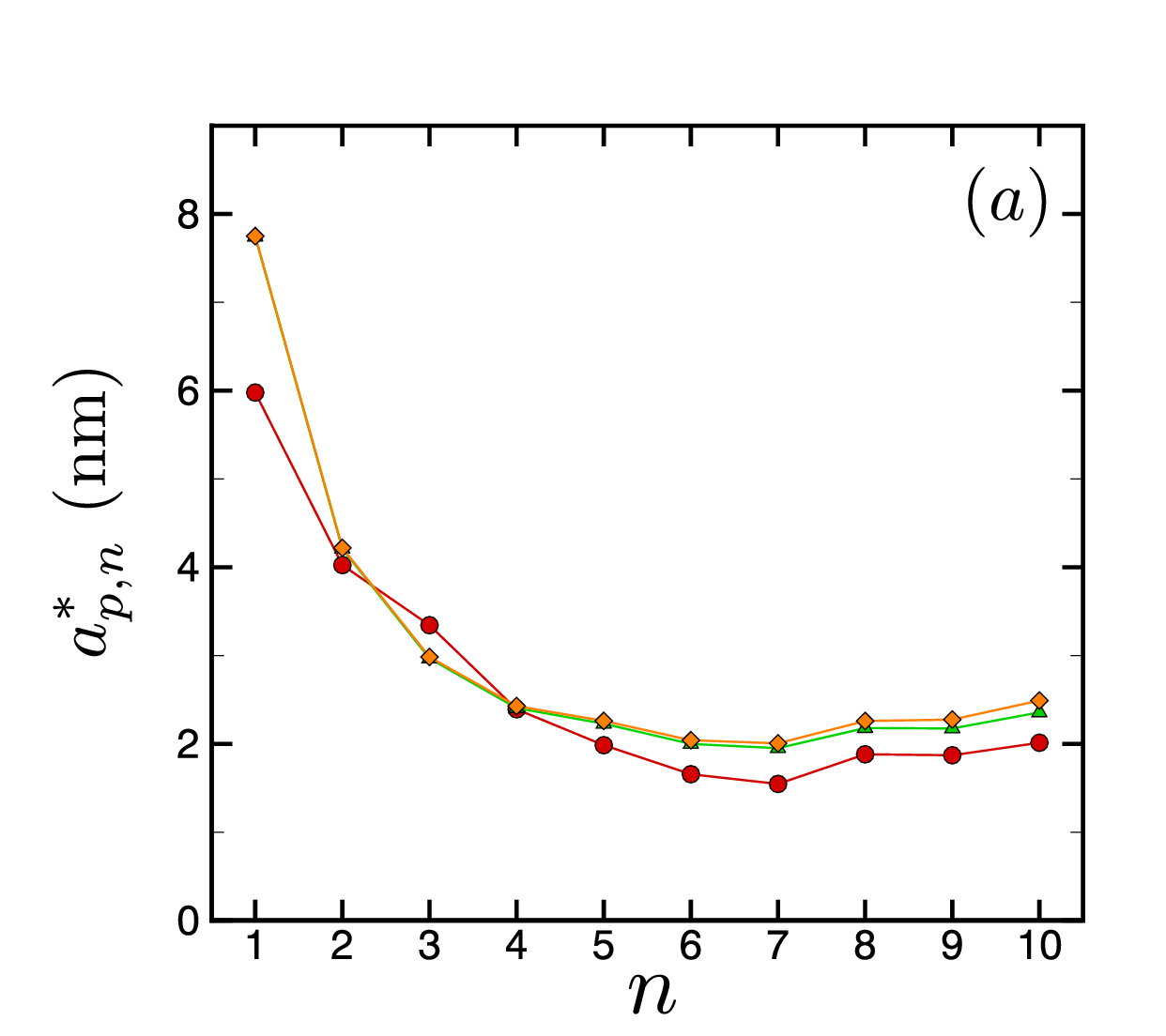}
\includegraphics[width=3.0in]{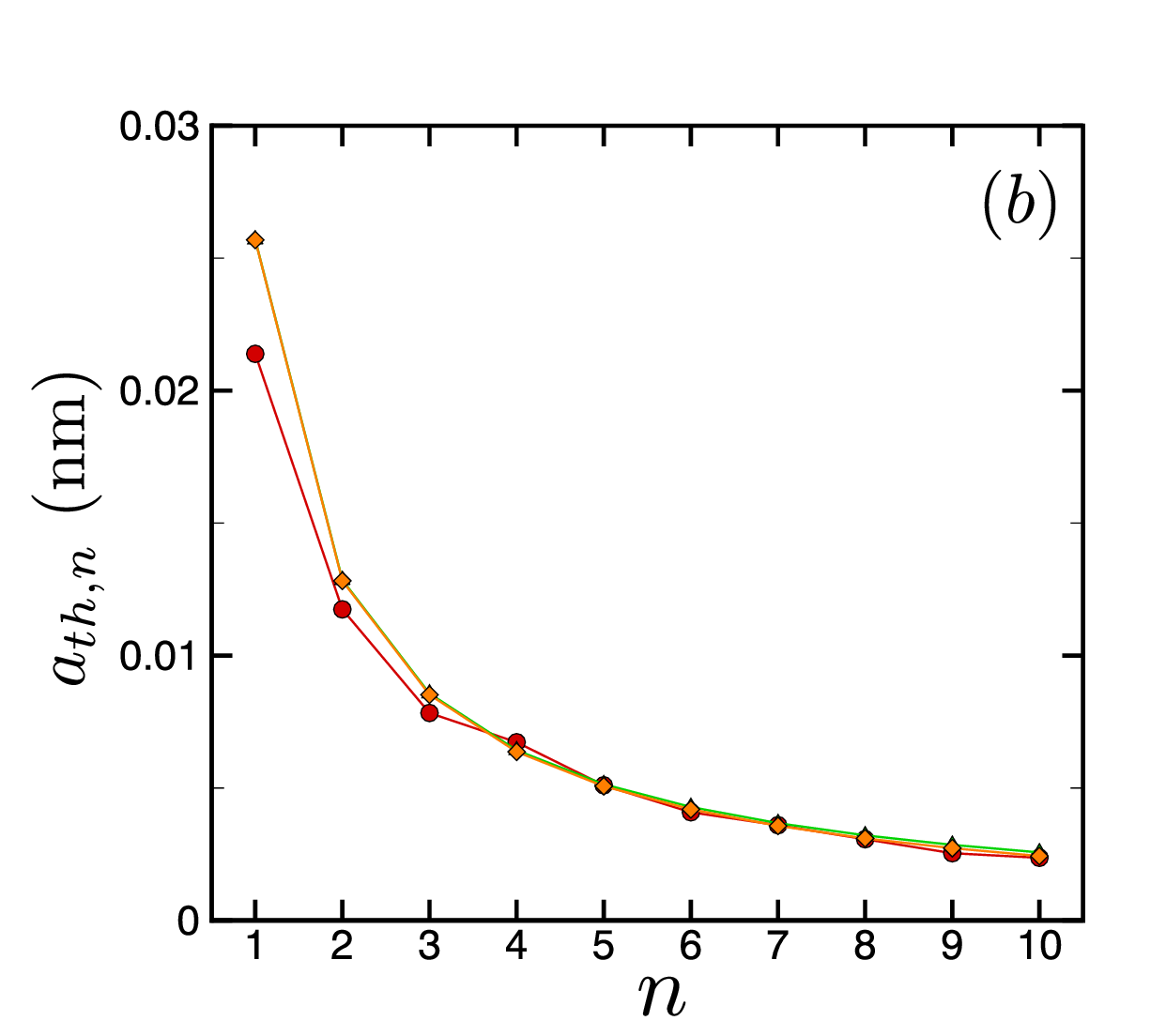}
\includegraphics[width=3.0in]{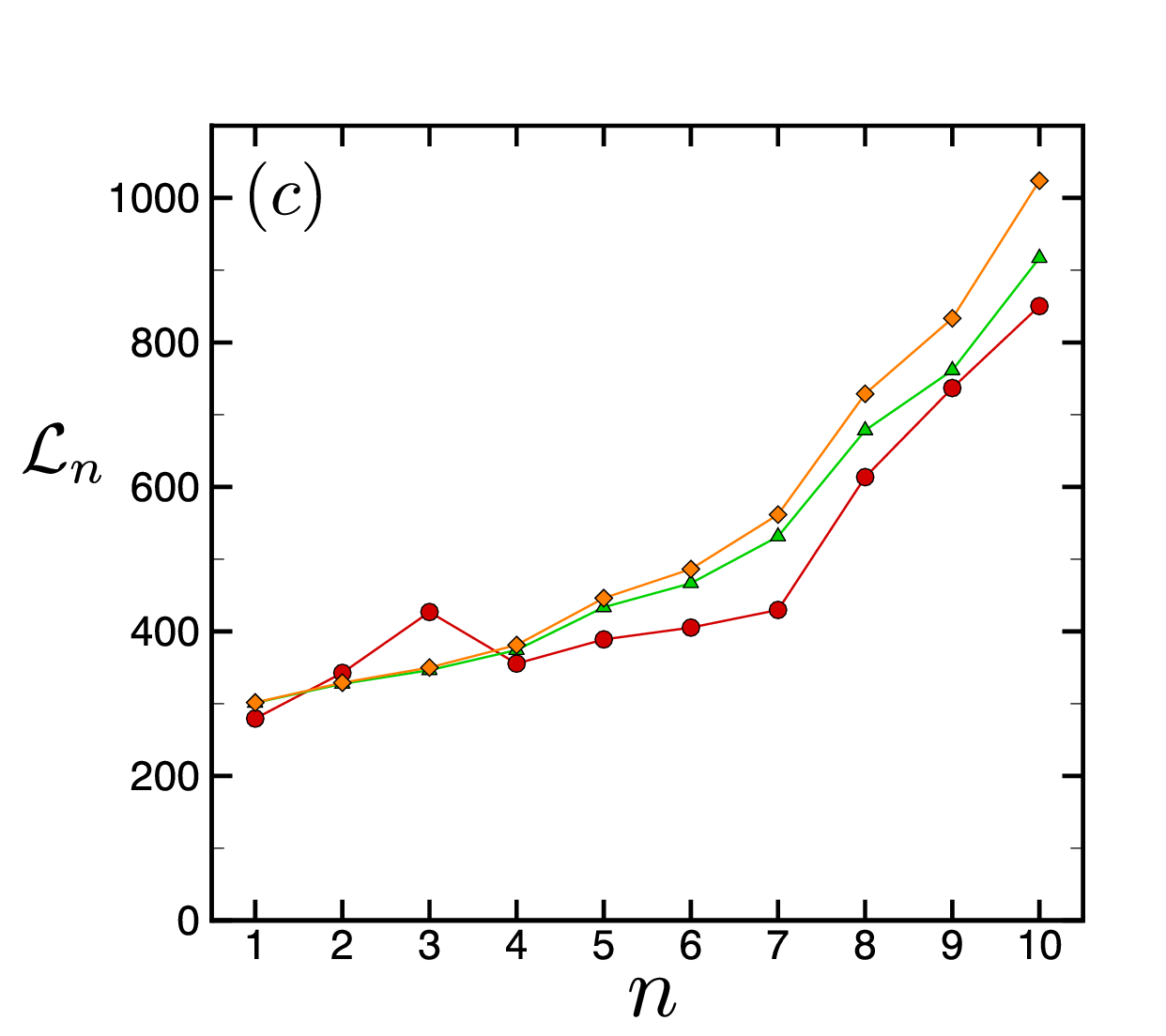}
\end{center}
\caption{Comparison between theory and experiment for the high tension beam in Table~\ref{table:beam} with $U\!=\!4233$. The variation of the ($a$) critical amplitude $a^*_{p,n}$, ($b$) the thermal amplitude $a_{th,n}$, and the ($c$) linear dynamic range $\mathcal{L}_n$. Circles (red) are measurements from experiment. Theoretical predictions are shown for a string (triangles, green) and a hinged beam (diamonds, orange).} 
\label{fig:beam-U4233}
\end{figure}

The comparison between theory and experiment for the high tension beam, beam 2, is shown in Fig.~\ref{fig:beam-U4233} using the same conventions as Fig.~\ref{fig:beam-U214}. For this very high tension case we only include the hinged result, and not the clamped result, since these two results are nearly indistinguishable for the results shown. In addition, for such a large value of $U$ it becomes very difficult to numerically evaluate the results for the clamped beam where the hinged beam results are determined using closed form expressions. 

Figure~\ref{fig:beam-U4233}(a) shows the variation of the critical RMS amplitude $a^*_{p,n}$. Due to the high intrinsic tension in the beam, its critical RMS amplitude is over twice that of beam 1. The agreement between the hinged description and the string description is very good for this case. This is due to the high tension, bending does not become significant until much higher mode numbers. Computing $n_c$ for $U\!=\!4233$ yields that the bending contribution would not equal the tension contribution until a mode number of $n_c \!\approx\! 30$. The deviation between the string and the hinged beam description remains small for $n \!\lesssim\! 8$. The string description is, in fact, slightly closer to the experimental values for the larger mode numbers. This deviation is very small and is most likely not significant.  The RMS amplitude of thermal motion is shown in Fig.~\ref{fig:beam-U4233}(b) and the agreement is very good for all of the results shown. The linear dynamic range is shown in Fig.~\ref{fig:beam-U4233}(c) where again the agreement is very good.

\section{Conclusion}
\label{section:conclusion}

We have explored the mode-dependent linear dynamic range of micro and nanoscale elastic beams that are currently used in a wide range of experiments. Most beams composed of silicon nitride are under significant intrinsic tension as a result of the fabrication process. When they are driven strongly, an additional stretching induced tension leads to a nonlinear response which can be described as a Duffing oscillator.

We have theoretically modeled this using the full Euler-Bernoulli beam theory with the inclusion of an intrinsic tension term. Although tractable, this approach does not yield analytical closed-form expressions which limits its ability to provide physical insights. However, in the presence of significant intrinsic tension we also explored two models that yield a much simpler description: a string model and a hinged-beam model. The string model provides immediate insight into the role of tension on the critical amplitude and is useful for cases where the tension is high and when one is not interested in the larger mode numbers. An estimate for how large a mode number can be used for the string model to be accurate within $\sim 5\%$ is given by $n_c$ in Fig.~\ref{fig:nc} using $\psi_n \!=\! 10$.

Lastly, the hinged-beam model uses the full beam equation with bending and tension and yields very useful closed-form analytical expressions. Importantly, the hinged beam results become more accurate with increasing mode number. The largest errors for the hinged model occur for the lowest mode numbers. However, these errors are small for $U \!\gtrsim\! 50$ as shown in Fig.~\ref{fig:acqoverh-hinged-comp}(a). For smaller values of the tension parameter, one should use the clamped beam results which are readily accessible when the tension is low. 

We anticipate that the findings presented here, and in particular the analytical expressions, will be of broad and immediate use for the determination, and use of, the linear dynamic range of multimodal beam oscillations for a wide range of conditions. Our results are quite general and it is straightforward to extend these ideas to more complex situations, geometries, and boundary conditions. For example, it would be interesting to explore the role of soft-clamped boundaries~\cite{golokolenov:2023} due to the undercut that is typical of many nanoscale beams. Although we have not focused upon the dissipation in the system, it would be insightful to explore the role of $Q_n$, including nonlinear dissipation~\cite{catilini:2021}, for different situations of interest~\cite{engelsen:2024} on the values of the critical amplitude $a_{c,n}$. 

\begin{acknowledgments}
\noindent M. R. Paul and N. W. Welles acknowledge support from the National Science Foundation (NSF) Grant No. CMMI-2001559. K. L. Ekinci and M. Ma acknowledge support from the NSF Grant Nos. CMMI-1934271 and CMMI-2001403.  
\end{acknowledgments}

\noindent \textbf{Conflict of Interest}

\noindent The authors have no conflict to disclose.

\noindent \textbf{Data Availability}

\noindent The data that supports the findings of this study are available from the corresponding author upon reasonable request.

\end{document}